\documentstyle[aps,prc,twocolumn,psfig]{revtex}

\begin{document}
\draft
\twocolumn

\newcommand{\bd}[1]{ \mbox{\boldmath $#1$}  }

\title{Breakup reactions of $^{11}$Li within a three-body model}

\author{E. Garrido} 
\address{Instituto de Estructura de la Materia,
CSIC, Serrano 123, E-28006 Madrid, Spain} 
\author{D.V.~Fedorov \and A.S.~Jensen} 
\address{Institute of Physics and Astronomy, Aarhus
University, DK-8000 Aarhus C, Denmark}

\date{\today}

\maketitle

\begin{abstract}
We use a three-body model to investigate breakup reactions of
$^{11}$Li (n+n+$^{9}$Li) on a light target. The interaction parameters
are constrained by known properties of the two-body subsystems, the
$^{11}$Li binding energy and fragmentation data. The remaining degrees
of freedom are discussed.  The projectile-target interactions are
described by phenomenological optical potentials. The model predicts
dependence on beam energy and target, differences between longitudinal
and transverse momentum distributions and provides absolute values for
all computed differential cross sections. We give an almost complete
series of observables and compare with corresponding
measurements. Remarkably good agreement is obtained. The relative
neutron-$^{9}$Li $p$-wave content is about 40\%. A $p$-resonance,
consistent with measurements at about 0.5 MeV of width about 0.4 MeV,
seems to be necessary. The widths of the momentum distributions are
insensitive to target and beam energy with a tendency to increase
towards lower energies. The transverse momentum distributions are
broader than the longitudinal due to the diffraction process. The
absolute values of the cross sections follow the neutron-target cross
sections and increase strongly for beam energies decreasing below 100
MeV/u.\\

\noindent
PACS number(s): 25.60.Gc, 21.60.Gx, 25.60-t, 21.45.+v
\end{abstract}

\section{Introduction}
Large efforts are devoted to investigate the properties of halo nuclei
\cite{han95,jon98}.  In particular two-neutron halo nuclei have
attracted a lot of attention with $^6$He (n+n+$^4$He) and $^{11}$Li
(n+n+$^{9}$Li) as the most prominent examples. These nuclei are also
Borromean three-body systems, where all two-body subsystems are
unbound \cite{zhu93,fed94}.  The two neutrons (the halo) are weakly
bound to an ordinary nucleus (the core). The halo is spatially
extended and the two neutrons have a high probability of being outside
the core. The core and halo degrees of freedom then approximately
decouple and three-body models provide a good description of such
systems.

The most detailed properties of halo nuclei are obtained by
measurements of fragment momentum distributions in breakup reactions
on stable targets
\cite{kob88,ann90,rii92,bla93,orr95,zin95,nil95,hum95,orr97,zin97,gei97,ale98,chu97}. The
projectile energy in these reactions is very large compared to the
energies of the intrinsic motion of the nucleons in the core, which in
turn is much larger than the binding energies of the spatially
extended halo particles. Such high-energy reactions are tempting to
describe in the sudden approximation where the three-body binding is
removed without disturbing the motion of the constituent
particles. The three halo particles continue their motion
independently without any further interaction. The resulting momentum
distributions then reflects the motion in the initial halo bound state
of the two neutrons and the core. Thus the unchanged initial
three-body wave function should describe the observed momentum
distributions. The Fourier transform of the wave function indeed
approximately reproduce the core momentum distribution
\cite{zhu93,zhu94,kor94,zhu95,gar96,gar97}. However, the calculated
neutron distributions are significantly broader than measured.

Improvements using Glauber theory are possible \cite{oga92,ber98}, but
the neutron distributions within this approach are only reported for
two-neutron halos in \cite{ber98}, where the two-neutron removal cross
sections and the related momentum distributions are nicely described.
However, an alternative is to improve the physically intuitive
geometric picture established by the successful sudden
approximation. The next step in such a description then amounts to
instantaneous removal of one halo particle (participant) while the
remaining two particles (spectators) continue to interact on the way
to detection.  This modification has only little influence on the core
momentum distribution but affects significantly the distribution of
the lighter neutron. Several authors suggested that this final state
interaction between the spectators plays an essential role, especially
when low-lying resonances are present \cite{zin95,kor94,bar93}. Indeed
computations then reproduce the measured momentum distributions
remarkably well when the final state interaction in a consistent
calculation is precisely the same as in the initial three-body wave
function \cite{gar96,gar97}. In addition the model also fairly well
describes the invariant mass spectra for the two-body system
consisting of the core and the remaining neutron.

This successful model requires, however, a better justification. The
sudden approximation assumes that the transition amplitude is
proportional to the overlap between initial and final state wave
functions and the momentum distributions are simply proportional to
the square of this overlap. The participant-target interactions,
implicitly used, are described as the schematic black sphere
scattering where only absorption is considered.  The obvious
improvement is to use the phenomenological optical model to describe
the interaction between the participant and the target
\cite{gar98,gar98b}. The qualitative improvement is then inclusion of
elastic scattering in addition to absorption.

At the high energies, where these models mostly are tested, neutron
elastic scattering is about three times smaller than neutron
absorption \cite{coo93}. However, the neutron momentum distribution is
much broader for elastic scattering than for absorption. The
contribution from neutron scattering is therefore important. Whether
the predictions would agree with measurements still remains to be
studied systematically. The first important step is to assume that the
optical model describes the interaction between the target and the
participant whereas the spectators remain undisturbed by the target
but still continue their motion under the influence of their mutual
interaction.  The complete three-body breakup reaction is then
described as a sum of these three independent contributions. We then
neglect processes where two or three halo particles interact
simultaneously with the target \cite{gar98,gar98b}.

The three-body model is strictly only valid for structureless
particles. The effects of extended density distributions for the
constituent particles are small when the major part of the three-body
wave function is outside the radii of all the three
particles. However, this requirement is usually not completely
fulfilled for halo nuclei. The model is then only accurate for the
outer part of the wave function. Furthermore, the spatial extensions
of the particles and the target allow geometric configurations where
more than one halo particle during the collision must get close to the
target. These configurations should be excluded in the process. For
one-nucleon halos this so-called shadowing effect is known to produce
smaller absolute cross sections and narrower momentum distributions
\cite{han96,esb96,hen96}. For two-neutron halos the shadowing effect
was included through profile functions in sophisticated Glauber
calculations of three-body fragmentation reactions \cite{ber98}. We
account for shadowing by excluding the participant wave function
inside spheres around the two spectators \cite{gar98,gar98b}.

The model then consists of an initial three-body halo state, reactions
caused by the participant-target optical potential, a final state with
two independent two-body systems, i.e. the two spectators and the
participant-target, and shadowing which excludes the wave function
within spheres around the spectators. The differential cross sections
are then products of the participant-target cross section and the
square of the overlap previously used in the pure sudden
approximation. The results from the successful sudden approximation
are essentially recovered, but the model now also, via the optical
potential, contains dependence on beam energy and target, distinction
between longitudinal and transverse momenta and also absolute values
of all the cross sections.

The purpose of this paper is to investigate the breakup reactions of
$^{11}$Li within the three-body model sketched above. We shall show
systematic computations of a number of observables and predict or
compare with measurements. The results are then all correlated as
arising from the same model with one set of parameters. In section II
we descibe the model and the method. In section III we compare our
results with the available experimental information and select an
interaction with corresponding shadowing parameters which best fits
the experimental data. Section IV presents predictions for a number of
observables including their energy dependence. Finally section V
contains a summary and the conclusions.

\section{Model and Method}

The spatially extended three-body halo collides with a relatively
small target at high energy. Then the probability that more than one
of the constituents interacts strongly with the target is small. The
differential cross section $d\sigma$ is then to a good approximation a
sum of three terms $d\sigma ^{(i)}$ each describing the independent
contribution to the process from the interaction between the target
and the halo particle $i$.  This is the assumption used in the
classical formulation for a weakly bound projectile \cite{ban67}. We
neglect the binding energy of the initial three-body bound state
compared to the high energy of the beam.  The reaction is then
described as three particles independently interacting with the target
as if each particle were free.

\begin{figure}[ht]
\centerline{\psfig{figure=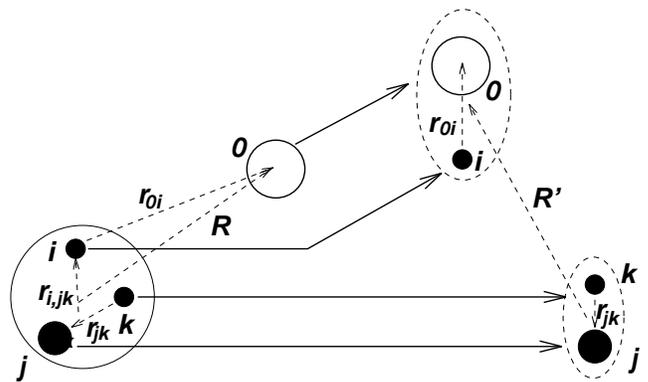,width=8.5cm,%
bbllx=6.5cm,bblly=6.5cm,bburx=15.2cm,bbury=21.2cm,angle=270}}
\vspace{0.2cm}
\caption[]{Sketch of the reaction and coordinates used. The target is
labelled by 0 and $\{i,j,k\}$ label the particles within the three--body
projectile. Compare with Eq.(\ref{e17}).}
\label{fig1}
\end{figure}

The process is described as removal of one particle $i$ (participant)
while the other two particles $j$ and $k$ (spectators) both survive
the reaction undisturbed (see Fig. \ref{fig1}.  The participant is
either absorbed (stripped) or elastically scattered (diffracted) by
the target.  The final state then consists of two independent
subsystems, i.e. the target plus participant and the two
spectators. The interaction in the final state between the two
spectators $j$ and $k$ must be the same as in the initial three-body
bound state. This consistency is previously only reported in
refs.\cite{gar96,gar97}.

We shall first show how to compute fragmentation cross sections of
two-neutron halo nuclei on light targets where the Coulomb interaction
is negligibly small. We then briefly describe how the wave functions
in the initial and final states are constructed. Finally we give
the various interactions used in the calculations.

\subsection{Cross sections}

The masses and the coordinates and their conjugate momenta are denoted
$m$, $\bd{r}$ and $\bd{p}$, respectively. The three halo particles and
the target are labeled by $\{i,j,k\}=\{1,2,3\}$ and $0$, respectively.
The relative coordinates $\bd{r}_{jk}, \bd{r}_{0i}, \bd{r}_{i,jk},
\bd{R}$ and $\bd{R}^{\prime}$, see Fig. \ref{fig1}, are defined by
\begin{eqnarray}
\bd{r}_{jk} & = &  \bd{r}_j - \bd{r}_k \;\; , \;\;\; 
\bd{r}_{0i}  =   \bd{r}_0 - \bd{r}_i \;\; ,  \nonumber \\
\bd{r}_{i,jk} & = & \bd{r}_i - 
         \frac{m_j \bd{r}_j + m_k \bd{r}_k}{m_j+m_k} \; , \nonumber \\
\bd{R} \equiv \bd{r}_{0,ijk} & = & \bd{r}_0 - 
     \frac{m_i \bd{r}_i + m_j \bd{r}_j + m_k \bd{r}_k}{m_i+m_j+m_k} \; ,
 \nonumber \\
\bd{R}^{\prime} \equiv \bd{r}_{0i,jk} & = & 
\frac{m_0 \bd{r}_0 + m_i \bd{r}_i}{m_0+m_i}
       - \frac{m_j \bd{r}_j + m_k \bd{r}_k}{m_j+m_k} \; . \label{e17} 
\end{eqnarray}
The corresponding conjugate momenta are analogously denoted by
$\bd{p}_{jk}, \bd{p}_{0i}, \bd{p}_{i,jk}, \bd{p}_{0,ijk}$ and
$\bd{p}_{0i,jk}$, i.e.
\begin{eqnarray} 
&& \bd{p}_{jk}  =  \frac{m_k \bd{p}_j - m_j \bd{p}_k}{m_k+m_j} \;\; , \;\;\; 
\bd{p}_{0i}  =\frac{m_i \bd{p}_0 - m_0 \bd{p}_i}{m_0+m_i} \; , \nonumber \\
&& \bd{p}_{i,jk} = \frac{(m_j+m_k)\bd{p}_i -m_i (\bd{p}_j+\bd{p}_k)}
                      {m_i+m_j+m_k}  \; , \nonumber \\ && \label{e2}
 \bd{P} \equiv \bd{p}_{0,ijk}  = \frac{(m_i+m_j+m_k) \bd{p}_0 - 
 m_0 (\bd{p}_i+\bd{p}_j+\bd{p}_k)}
                            {m_0+m_i+m_j+m_k}  \; , \nonumber \\
&& \bd{p}_{0i,jk} =  \frac{(m_j+m_k)(\bd{p}_0 + \bd{p}_i)
                       -(m_0+m_i)(\bd{p}_j + \bd{p}_k)}
                        {m_0+m_i+m_j+m_k}  . \label{e21}
\end{eqnarray}
These momenta are related in the same way for the final states.  We
use primes to indicate the final state and may therefore add primes on
all momenta in Eq.(\ref{e21}).  In particular $\bd{p}_{jk}^\prime$ and
$\bd{p}_{0i}^\prime$ are the relative two-body momenta and
$\bd{P}^\prime \equiv \bd{p}_{0i,jk}^\prime$ is the conjugate of
$\bd{R}^{\prime}$ in the final state.

We can now easily express the individual momenta in terms of three
relative momenta and one absolute momentum, for example
$\bd{p}_{jk},\bd{p}_{0i},\bd{p}_{0,ijk}$ and $\bd{p}_{i} + \bd{p}_{j}
+ \bd{p}_{k}$. This would be natural for the initial state whereas it
is more convenient to use the beam momentum and
$\bd{p}_{jk}^\prime,\bd{p}_{0i}^\prime,\bd{P}^\prime$ in the final
state. We shall work in the center of mass system of the projectile
where the beam momentum is $-\bd{p}_{0}$ and $\bd{p}_{i} + \bd{p}_{j}
+ \bd{p}_{k} = 0$. Using total momentum conservation,
$\bd{p}_{0}^\prime = \bd{p}_{0} - \bd{p}_{i}^\prime -
\bd{p}_{j}^\prime - \bd{p}_{k}^\prime$, we obtain the individual
momenta in the final state as
\begin{eqnarray}
\bd{p}^\prime_j&=&\bd{p}^\prime_{jk} - \frac{m_j \bd{P}^\prime}{m_j+m_k} 
                + \frac{m_j \bd{p}_0}{m_0+m_i+m_j+m_k}  \; ,
\nonumber \\
\bd{p}^\prime_k&=&-\bd{p}^\prime_{jk} - \frac{m_k \bd{P}^\prime}{m_j+m_k} 
                + \frac{m_k \bd{p}_0}{m_0+m_i+m_j+m_k}  \; ,
\nonumber \\
\bd{p}^\prime_i&=&-\bd{p}^\prime_{0i} + \frac{m_i \bd{P}^\prime}{m_0+m_i} 
                + \frac{m_i \bd{p}_0}{m_0+m_i+m_j+m_k} \; ,
\nonumber \\ \label{e23}
\bd{p}^\prime_0&=&\bd{p}^\prime_{0i} + \frac{m_0 \bd{P}^\prime}{m_0+m_i} 
                + \frac{m_0 \bd{p}_0}{m_0+m_i+m_j+m_k} \; .
\end{eqnarray}

The process is described as a participant-target reaction and two
undisturbed spectators, i.e. momentum conservation dictates that
\begin{equation} \label{e25}
\bd{p}^\prime_{j}+ \bd{p}^\prime_{k} =  \bd{p}_{j} + \bd{p}_{k} \;\; , \;\;
\bd{p}^\prime_{0}+ \bd{p}^\prime_{i} =  \bd{p}_{0} + \bd{p}_{i} \; .
\end{equation}
In the rest frame of the projectile $\bd{p}_{j} + \bd{p}_{k} = -
\bd{p}_{i}$ and $\bd{p}_{i} = \bd{p}_{i,jk}$. Then Eq.(\ref{e25}) can
be rewritten as
\begin{eqnarray} \label{e3}
\bd{p}_{i,jk}&=& \bd{P}^\prime - \frac{m_j+m_k}{m_i+m_j+m_k+m_0} \bd{p}_0 \; , 
 \nonumber \\    \bd{p}_{0i}&=& \frac{m_i+m_j+m_k}{m_i+m_j+m_k+m_0} \bd{p}_0
 - \frac{m_0}{m_i+m_0} \bd{P}^\prime \; .
\end{eqnarray}

Let us now first consider elastic scattering of the participant $i$ by
the light target.  We denote the initial three-body projectile wave
function by $\Psi^{(JM)}(\bd{r}_{jk},\bd{r}_{i,jk})$, where $J$ and
$M$ are the total angular momentum and its projection on the
$z$-axis. The final state outgoing distorted wave functions of the
independent participant-target and spectator two-body subsystems are
respectively $\phi _{ \bd{p}_{0i}^{\prime }\Sigma
^{\prime}_i}^{(0i+)}$ and $\phi _{\bd{p}_{jk}^{\prime }s_{jk}
\Sigma_{jk}}^{(jk+)}$, where $\bd{s}_{jk}= \bd{s}_{j}+\bd{s}_{k}$,
$\Sigma_{jk} = \Sigma_{j} + \Sigma_{k}$ and $s_{i}$ and $\Sigma_i$ are
spin and projection quantum numbers of particle $i$. The corresponding
spin functions are $\chi _{s_{i}\Sigma^{\prime } _{i}}$, $\chi
_{s_{jk}\Sigma _{jk}}$, where we for convenience assume a zero spin
target. The participant-target interaction is $V_{0i}$, where we only
consider a light target. A possible Coulomb interaction can then be
neglected.

The transition amplitude $T^{(i)}$ of the process, where particle $i$
is elastically scattered by the target while particles $j$ and $k$ are
undisturbed, is most conveniently computed in the center of mass
system of the three-body projectile. Using momentum conservation we
obtain
\begin{equation}
T^{(i)}= \langle \phi_{ \bd{p}_{0i}^\prime  \Sigma^\prime_i}^{(0i+)} 
\phi_{ \bd{p}_{jk}^\prime s_{jk} \Sigma_{jk}}^{(jk+)} e^{i\bd{P}^\prime
\bd{R}^\prime }|V_{0i}|\Psi^{(JM)}
e^{i\bd{P}\bd{R}} \rangle \; ,
\label{eq1}
\end{equation}
This amplitude basically factorizes into particant-target elastic
scattering transition amplitude $T^{(0i)}_{\Sigma_i
\Sigma^{\prime}_i}$ and the overlap between initial and final states
of the spectators $M_{s_{jk} \Sigma_{jk} \Sigma_i}^{(JM)}$, i.e.
\begin{eqnarray}
& T^{(i)}=\sum_{\Sigma_i} T^{(0i)}_{\Sigma_i \Sigma^\prime_i}
M_{s_{jk} \Sigma_{jk} \Sigma_i}^{(JM)} \; , \\
& T^{(0i)}_{\Sigma_i \Sigma^{\prime}_i} =
\langle \phi_{\bd{p}_{0i}^\prime \Sigma^\prime_i}^{(0i+)}|V_{0i}|
e^{i\bd{p}_{0i}\bd{r}_{0i}}
  \chi_{s_i \Sigma_i} \rangle 
\label{eq2}  \; , \\
& M_{s_{jk} \Sigma_{jk} \Sigma_i}^{(JM)} = \langle
\phi _{\bd{p}_{jk}^{\prime }s_{jk} \Sigma_{jk}}^{(jk+)} e^{i
\bd{p}_{i,jk}\bd{r}_{i,jk}}\chi _{s_{i}\Sigma _{i}}|\Psi^{(JM)}\rangle \; .
\label{eq3}
\end{eqnarray}
In breakup computations, where only absorption is included in the
sudden approximation, the transition amplitude is proportional to the
overlap $M_{s_{jk} \Sigma_{jk} \Sigma_i}^{(JM)}$, see
\cite{gar96,gar97}. This previous approximation is therefore still
valid provided the elastic scattering process can be neglected. If
furthermore the final state interaction between the two spectators is
neglected the overlap in Eq.(\ref{eq3}) reduces to the Fourier
transform of the projectile wave function $\Psi^{(JM)}$. This is the
approximation used in the first attempt to understand these
fragmentation reactions \cite{zhu93,zhu94,kor94,zhu95}.

The differential diffraction (elastic scattering) cross section is
then given by
\begin{eqnarray} \label{e7}
d^9\sigma_{el}^{(i)} & = & \frac{2\pi}{\hbar} \frac{1}{v} 
\frac{\delta (E_{0i}^{\prime }-E_{0i})}{2J+1} \nonumber \\ 
& \times & \sum_{M s_{jk} \Sigma_{jk} \Sigma^\prime_i} \left| 
\sum_{\Sigma_i} T^{(0i)}_{\Sigma_i \Sigma^\prime_i}
M_{s_{jk} \Sigma_{jk} \Sigma_i}^{(JM)}
\right|^2 d\nu _{f}^{(i)} \; ,
\end{eqnarray}
where $v=p_0/m_0$ is the velocity of the target seen from the
projectile rest frame and $E_{0i}=p_{0i}^2/2\mu_{0i}$ and
$E^\prime_{0i}=p_{0i}^{\prime 2}/2\mu_{0i}$ are the relative energies
of particle $i$ and the target in the initial and final states. We use
here the non-relativistic expressions, since the optical model is
non-relativistic although obtained through a relativistic procedure,
see the discussion later in the paper. The reduced mass of particle
$i$ and the target is here denoted $\mu_{0i}$. Then energy
conservation demands that $|\bd{p}_{0i}|=|\bd{p}_{0i}^\prime|$. The
density of final states $d\nu _{f}^{(i)}$ is given by
\begin{equation}
d\nu _{f}^{(i)} =  
\frac{d^{3}{\bd{p}}_{0i}^{\prime }}{(2\pi
\hbar )^{3}}\frac{d^{3}{\bd{p}}_{jk}^{\prime }}{(2\pi \hbar
)^{3}}\frac{d^{3}
{\bd{P}}^{\prime }}{(2\pi \hbar )^{3}} \; . \label{pha}
\end{equation}

The differential diffraction cross section in Eq.(\ref{e7}) can now
also be rewritten in factorized form. When the participant $i$ has
spin 0 or 1/2 and the target has spin 0, we get, as shown in Appendix
A, the expression
\begin{eqnarray}
&&  \frac{d^9\sigma _{el}^{(i)}(\bd{P}^{\prime },
\bd{p}_{jk}^{\prime },\bd{p}_{0i}^{\prime })}
{ d\bd{P}^{\prime } d\bd{p}_{jk}^{\prime } d\bd{p}_{0i}^{\prime } }
  =   \nonumber \\ && \; \;  \frac{d^3\sigma _{el}^{(0i)}(\bd{p}_{0i}
  \rightarrow  \bd{p}_{0i}^{\prime})}
 {d\bd{p}_{0i}^{\prime }} \; 
 |M_s(\bd{p}_{i,jk}, \bd{p}_{jk}^{\prime })|^2 \; ,  \label{eq5}
\end{eqnarray}
where the first factor is the differential cross section for the
participant-target elastic scattering process, 
\begin{eqnarray}
 \frac{d^3\sigma _{el}^{(0i)}(\bd{p}_{0i}
  \rightarrow \bd{p}_{0i}^{\prime})} {d\bd{p}_{0i}^{\prime }}
& = &\frac{1}{v}\frac{2\pi }{\hbar } 
 \frac{\delta (E_{0i}^{\prime }-E_{0i})}{(2\pi \hbar )^{3}}
 \nonumber \\ & \times &\frac{1}{2s_i+1} \sum_{\Sigma_i\Sigma^{\prime}_i}
|T^{(0i)}_{\Sigma_i \Sigma^{\prime}_i}|^2 \; , \label{el1}
\end{eqnarray}
and the second factor is the overlap matrix element used in the
original formulation of the sudden approximation for absorption
\cite{gar96,gar97}
\begin{equation}
   |M_s(\bd{p}_{i,jk}, \bd{p}_{jk}^{\prime })|^2 \equiv 
\frac{1}{2 J+1} \sum_{M s_{jk}\Sigma_{jk}\Sigma_i }
 |M_{s_{jk} \Sigma_{jk} \Sigma_i}^{(JM)}|^{2} \; ,
\label{sudden}
\end{equation}
which is normalized to one, i.e. gives unity after integration over
$\bd{p}_{i,jk}$ and $\bd{p}_{jk}^{\prime }$ or equivalently, by use of
Eq.(\ref{e3}), over $\bd{P}^{\prime}$ and $\bd{p}_{jk}^{\prime }$.

We also consider the other process, where the participant in the sense
of the optical model is absorbed by the target. We obtain analogously
the corresponding differential absorption (stripping) cross section in
the same factorized form
\begin{eqnarray} \label{eq6}
 \frac{d^6\sigma _{abs}^{(i)}(\bd{P}^{\prime },\bd{p}_{jk}^{\prime })}
{ d\bd{P}^{\prime } d\bd{p}_{jk}^{\prime } }
=   \sigma _{abs}^{(0i)}(p_{0i}) \; 
 |M_s(\bd{p}_{i,jk}, \bd{p}_{jk}^{\prime })|^2 \; , 
\end{eqnarray}
where $\sigma _{abs}^{(0i)}$ is the participant-target absorption
cross section. The nine-dimensional differential cross section is now
reduced to six, since the absorped or stripped particle inherently is
of no interest in the optical model description. 

It is conceptually important to realize that the factorizations in
Eqs.(\ref{eq5}) and (\ref{eq6}) are incomplete, since $\bd{p}_{0i}$ in
the participant-target cross sections is related to $\bd{P}^{\prime}$
or ($\bd{p}_{i,jk}$) in Eq.(\ref{sudden}) via the momentum
conservation in Eq.(\ref{e3}).  For high energy reactions the
factorization is in practice fairly accurate for two reasons.  First
the range of $p_{0i}$-values is limited to an interval around
$\frac{m_i}{m_0+m_i}p_0$ determined by the size of the relatively
small momenta in the motion of the particles within the projectile,
see Eq.(\ref{e21}). Second the factor arising from the
participant-target cross section depends only weakly on energy for the
large beam energies corresponding to this rather small range of
$p_{0i}$-values.

Computations of the fragmentation cross sections are now essentially
reduced to computations of the overlap matrix element $M_{s_{jk}
\Sigma_{jk},\Sigma_i}^{(JM)}$ from \cite{gar97} (modified to account
for shadowing by omission of the unwanted cnofiguartions) and the
two-body elastic and absorption cross sections determined by the
optical model phase shifts. A target with zero spin, e.g. an even-even
nucleus like $^{12}$C, and a neutron with spin 1/2 as the participant
particle is of particular interest for halo nuclei. We then have
\cite{sit71}
\begin{eqnarray}
 \frac{d^3\sigma _{el}^{(0i)}(\bd{p}_{0i}
  \rightarrow \bd{p}_{0i}^{\prime})} {d\bd{p}_{0i}^{\prime }} & = & 
\frac{\delta (E_{0i}^{\prime }-E_{0i})}{\mu_{0i}^{2} v} 
\nonumber \\ \label{el2}
& \times & \left( |g(p_{0i}, \theta)|^2 + |h(p_{0i}, \theta)|^2 \right) \; ,
\end{eqnarray}
\begin{eqnarray}
 \sigma _{abs}^{(0i)}(p_{0i}) & = & 
\frac{\pi}{p_{0i}^2} \sum_{\ell=0}^{\infty}
\Bigl[ (2\ell +1)   \nonumber \\
& - &(\ell+1)  \left| e^{2i\delta_\ell^{(\ell+1/2)}} \right|^2   -
\ell \left| e^{2i\delta_\ell^{(\ell-1/2)}} \right|^2 \Bigr] \; ,
\end{eqnarray}
where $\theta$ is the angle between $\bd{p}_{0i}$ and
$\bd{p}_{0i}^{\prime}$ and $\delta_\ell^{(j)}$ is the phase shift of
the partial wave $\ell$ when the total angular momentum is $j$. The
functions $g$ and $h$ are given by
\begin{eqnarray}
g(p_{0i}, \theta)=\frac{1}{2 i p_{0i}} \sum_{\ell=0}^{\infty} 
\Bigl[ (\ell +1) (e^{2i\delta_\ell^{(\ell+1/2)}}-1)  \nonumber \\
 + \; \ell \; (e^{2 i \delta_\ell^{(\ell-1/2)}}-1) \Bigr] \; 
P_\ell(\cos \theta)  \; , \label{gfu}  
\end{eqnarray}
\begin{eqnarray} 
h(p_{0i}, \theta)= \frac{1}{2 p_{0i}} \sum_{\ell=1}^{\infty}
\left( e^{2i\delta_\ell^{(\ell+1/2)}}-e^{2i\delta_\ell^{(\ell-1/2)}} \right)
 \nonumber \\ \label{hfu} \times
\sin \theta \frac{d}{d(\cos \theta)} P_\ell(\cos \theta) \; ,
\end{eqnarray}
where $P_\ell$ is the $\ell$'th Legendre polynomial.

The two contributions (stripping and diffraction or absorption and
scattering) arising from the interaction $V_{0i}$ to any measurable
cross section is now obtained by integration over the unobserved
momenta in Eqs.(\ref{eq5}) and (\ref{eq6}).  The total cross section
is given by the sum of both contributions and the weight of each of
them is directly dictated by the optical potential. We shall compute
individual as well as relative momentum distributions in the final
state both along $\bd{p}_0$ (longitundinal) and perpendicular to
$\bd{p}_0$ (transverse). 

In addition to momentum distributions we shall also compute other
observables like the invariant mass spectrum of the two particles in
the final state. This mass is invariant under Lorentz transformations
and therefore independent of coordinate system. In particular, in the
rest system for the particles $j$ and $k$ the invariant mass reduces
for the relevant small energies to the non-relativistic kinetic energy
$E_{jk} = E_{jk}^{\prime} = p_{jk}^{\prime 2}/2\mu_{jk}$ of this
two-body system. Then the momentum variable $p_{jk}^{\prime }$ can be
substituted by $E_{jk}$ by use of $d E_{jk} = p_{jk}^{\prime } d
p_{jk}^{\prime }/\mu_{jk}$.

Another recently investigated observable is the angular distribution
of the relative momentum between the spectators $j$ and $k$ in a
coordinate system with the $z$-axis along the center of mass of their
total momentum in the final state. The decisive variable is then the
angle between $\bd{p}_{jk}^\prime$ and $\bd{p}_{j}^\prime +
\bd{p}_{k}^\prime = - \bd{p}_{i} = - \bd{p}_{i,jk}$, where the latter
vector is given in Eq.(\ref{e3}). To compute this angular distribution
we express the differential cross sections as functions of
$-\bd{p}_{i,jk}$ and $\bd{p}_{jk}^\prime$ and integrate over all the
variables except the angle between these two vectors. 

The necessary integrations require variable changes in the description
of the final state. The expressions in Eq.(\ref{e23}) provide useful
identities for this purpose, i.e.
\begin{eqnarray}
\frac{d^9\sigma _{el}^{(i)}(\bd{P}^{\prime },
\bd{p}_{jk}^{\prime },\bd{p}_{0i}^{\prime })}
{ d\bd{P}^{\prime } d\bd{p}_{jk}^{\prime } d\bd{p}_{0i}^{\prime } } =
\frac{d^9\sigma _{el}^{(i)}(\bd{P}^{\prime },
\bd{p}_{j}^{\prime },\bd{p}_{0i}^{\prime })}
{ d\bd{P}^{\prime } d\bd{p}_{j}^{\prime } d\bd{p}_{0i}^{\prime } } = 
\nonumber \\ \frac{d^9\sigma _{el}^{(i)}(\bd{P}^{\prime },
\bd{p}_{k}^{\prime },\bd{p}_{0i}^{\prime })}
{ d\bd{P}^{\prime } d\bd{p}_{k}^{\prime } d\bd{p}_{0i}^{\prime } } =
\frac{d^9\sigma _{el}^{(i)}(\bd{P}^{\prime },
\bd{p}_{jk}^{\prime },\bd{p}_{i}^{\prime })}
{ d\bd{P}^{\prime } d\bd{p}_{jk}^{\prime } d\bd{p}_{i}^{\prime } } \; , \\
\frac{d^6\sigma _{abs}^{(i)}(\bd{P}^{\prime },\bd{p}_{jk}^{\prime })}
{ d\bd{P}^{\prime } d\bd{p}_{jk}^{\prime } } =
\frac{d^6\sigma _{abs}^{(i)}(\bd{P}^{\prime },\bd{p}_{j}^{\prime })}
{ d\bd{P}^{\prime } d\bd{p}_{j}^{\prime } }=
\frac{d^6\sigma _{abs}^{(i)}(\bd{P}^{\prime },\bd{p}_{k}^{\prime })}
{ d\bd{P}^{\prime } d\bd{p}_{k}^{\prime } } 
\end{eqnarray}

We have so far only considered the two contributions coming from the
stripping and the diffraction of particle $i$ via the interaction
$V_{0i}$. However, the reaction in question may also be a result of
the other interactions $V_{0j}$ or $V_{0k}$. These cross sections are
then simply added. Possible interference terms are neglected, since
they in any case are small for spatially extended projectiles. From
Eqs.(\ref{eq5}) and (\ref{eq6}) we expect that the absolute values all
are of the same order of magnitude as the corresponding
participant-target cross sections.

\subsection{Wave functions and shadowing}

The initial three-body wave function $\Psi^{(JM)}$ of the projectile
is obtained by solving the Faddeev equations in coordinate space
\cite{fed94}.  We use the three sets of hyperspherical coordinates
($\rho, \Omega_i$), $\Omega_i = \{\alpha_i, \Omega_{xi}, \Omega_{yi}\}$,
where each $i$ is related to a given Jacobi system, see
\cite{zhu93,fed94}.  Then $\Psi^{(JM)}$ is a sum of the three Faddeev
components, which in turn for each hyperradius $\rho$ are expanded in
a complete set of generalized angular functions
$\Phi^{(i)}_{n}(\rho,\Omega_i)$
\begin{equation}
\Psi^{J M}= \frac {1}{\rho^{5/2}}
  \sum_n f_n(\rho)
\sum_{i=1}^3 \Phi^{(i)}_{n}(\rho ,\Omega_i) \; ,
\label{tot}
\end{equation}
where $\rho^{-5/2}$ is related to the volume element $\rho^5 d\rho
d\Omega_i$ with the angular part $d\Omega_i=\sin^2\alpha_i
\cos^2\alpha_i d\alpha_i d\Omega_{xi} d\Omega_{yi}$.

These angular wave functions satisfy the angular part of the three
Faddeev equations, i.e.
\begin{eqnarray}
 {\hbar^2 \over 2m}\frac{1}{\rho^2}\hat\Lambda^2 \Phi^{(i)}_{n}
 +V_{jk} (\Phi^{(i)}_{n}+\Phi^{(j)}_{n} + \Phi^{(k)}_{n}) 
 \nonumber \\
\equiv {\hbar^2 \over
2m}\frac{1}{\rho^2} \lambda_n(\rho) \Phi^{(i)}_{n}  \; ,
\label{ang}
\end{eqnarray}
where $\{i,j,k\}$ is a cyclic permutation of $\{1,2,3\}$, $m$ is a
normalization mass, $V_{jk}$ is the two-body interaction between
particles $j$ and $k$ and $\hat\Lambda^2$ is the $\rho$-independent
part of the kinetic energy operator. The analytic expressions for
$\hat{\Lambda}^2$ and the kinetic energy operator can, for instance, be
found in \cite{fed94}.

The radial expansion coefficients $f_n(\rho)$ are obtained from
the coupled set of ``radial'' differential equations \cite{fed94}
\begin{eqnarray} \label{rad}
   \left(-\frac{\rm d ^2}{\rm d \rho^2}
   -{2m(E-V_3(\rho)) \over\hbar^2}+ 
   \frac{ \lambda_n(\rho) }{\rho^2} + \frac{15}{4\rho^2}
 - Q_{n n} \right)  \nonumber \\ \times  f_n(\rho)
  = \sum_{n' \neq n}   \left(
   2P_{n n'}{\rm d \over\rm d \rho}
   +Q_{n n'}
   \right)f_{n'}(\rho)  \; ,
\end{eqnarray}
where $V_3$ is an anticipated three-body potential and the functions
$P$ and $Q$ are defined as the angular integrals
\begin{eqnarray}
   P_{n n'}(\rho)\equiv \sum_{i,j=1}^{3}
   \int d\Omega \Phi_n^{(i)\ast}(\rho,\Omega)
   {\partial\over\partial\rho}\Phi_{n'}^{(j)}(\rho,\Omega)  \; , \\
   Q_{n n'}(\rho)\equiv \sum_{i,j=1}^{3}
   \int d\Omega \Phi_n^{(i)\ast}(\rho,\Omega)
   {\partial^2\over\partial\rho^2}\Phi_{n'}^{(j)}(\rho,\Omega)  \; .
\end{eqnarray}

The continuum wave function $\phi _{{\bd p}_{jk}^{\prime }s_{jk}
\Sigma_{jk}}^{(jk+)}$ describing the two-body spectator system in the
final state is expanded in partial waves \cite{gar97}
\begin{eqnarray}
\phi _{{\bd p}_{jk}^{\prime }s_{jk} \Sigma_{jk}}^{(jk+)}
 =  \sqrt{\frac{2}{\pi}} \; 
\frac{1}{p^\prime_{jk} r_{jk}} \sum_{j_{jk} \ell_{jk} m_{jk}}
u_{\ell_{jk} s_{jk}}^{j_{jk}}(p^\prime_{jk}, r_{jk}) \nonumber \\ \times \;
{\cal Y}_{j_{jk} \ell_{jk} s_{jk}}^{m_{jk}*}(\Omega_{r_{jk}}) 
  \sum_{m_{\ell_{jk}}=-\ell_{jk}}^{\ell_{jk}}
i^{\ell_{jk}} Y_{\ell_{jk} m_{\ell_{jk}}} (\Omega_{p^\prime_{jk}})
 \nonumber \\ \times \;
\langle \ell_{jk} m_{\ell_{jk}}  s_{jk} \Sigma_{jk} | j_{jk} m_{jk}
\rangle  
 \; ,
\label{2beq}
\end{eqnarray}
where $\langle \; | \; \rangle$ is a Clebsch-Gordon coefficient,
$Y_{\ell m_{\ell}}$ is the spherical harmonic and ${\cal Y}_{j \ell
s}^{m*}(\Omega_{r})$ is the angular wave function obtained by coupling
orbital $\ell$ and spin $s$ to the total angular momentum and
projection $j$ and $m$. The distorted radial wave functions
$u_{\ell_{jk} s_{jk}}^{j_{jk}}(p^\prime_{jk}, r_{jk})$ are obtained by
solving the Schr\"{o}dinger equation with the appropriate two-body
potential. When this interaction between the two spectators in the
final state is neglected the expansion in Eq.(\ref{2beq}) reduces to
the usual expansion of plane waves in terms of spherical Bessel
functions.

The participant-target interaction is described by the optical
potential while the spectators remain unaffected. The finite extension
of the projectile and the target therefore in addition requires
exclusion of configurations where the spectators pass the target too
close to the participant. This corresponds to black sphere models
describing the spectator-target interactions. This so-called shadowing
strictly requires exclusion of the initial projectile wave function in
an infinitely long cylinder with the axis along the motion of the
participant. However, such a cylinder depends on the dynamics of the
reaction and omission of these events would be technically difficult
in large scale systematic computations. Instead we approximate the
shadowing by excluding spheres of the wave function where the
participant is close to the spectators. This is much simpler and has
also the appealing feature that major parts of the three-body wave
function describing configurations, where the spatially extended
particles are inside the radii of each other, simultaneously are
excluded. Then the contribution from densities, where halo and core
nucleons overlap, decreases and consequently the possible uncertainty
due to the treatment of the Pauli principle must be diminished or
perhaps completely eliminated \cite{gar97b}.

Thus we account for the shadowing effect by substituting zero for the
initial three-body wave function $\Psi^{(JM)}$ when the distances
between participant $i$ and the two spectators $j$ and $k$ are smaller
than the shadowing parameters $r_{nc}^{(ij)}$ and $r_{nn}^{(ik)}$,
respectively. Here we indicated that one neutron reacts with the
target while the core and the other neutron are spectators.  Instead
of this sharp cutoff a smooth function, varying from zero at small
distances to one at large distances, could easily be used to eliminate
the unwanted geometric configurations. The cutoff radii are in any
case related to the sizes of target and spectators. With the core or
nucleon as spectators a reasonable parametrization could then be
\begin{equation} \label{e29}
r_{nc} = r_0 \sqrt{ A_t^{2/3} + A_c^{2/3}} \;\; , \;\;
r_{nn} = \sqrt{ r_0^2 A_t^{2/3} + R_N^2} \; ,
\end{equation}
where we for simplicity omitted the superscripts $ij$ and $ik$.  The
mass numbers of target and core are $A_t$ and $A_c$ ($A_c=9$ for
$^{11}$Li) and $R_N \approx 1$ fm is the sharp cutoff radius of the
nucleon. We choose the parameter $r_0 \approx 1.26$ fm, which is
adjusted to reproduce the few reported absolute values of two-neutron
removal cross sections. This somewhat large value of $r_0$ may reflect
a parametrization where the range of the nuclear interaction is
included. The shadowing effect substantially reduces the absolute
values of the cross sections \cite{ber98,gar98}.  This sensitivity in
addition to the approximate nature of the treatment of shadowing
indicates that the predicted absolute values are less accurate than
desired. On the other hand the shapes of the distributions are fairly
insensitive to the shadowing parameters although the effects are
significant at the present level of accuracy.

\subsection{Interaction parameters}

The model described in the previous subsections needs specifications
of the interactions appropriate for the process under
investigation. We shall here focus on the fragmentation of a $^{11}$Li
($^{9}$Li + n + n) projectile on a carbon target and divide the
parameters into five sets, i.e. (i) the four two-body interactions
between neutron-neutron, (ii) neutron-$^{9}$Li, (iii) neutron-carbon
and (iv) $^{9}$Li-carbon two-body systems and (v) the three-body
interaction for the $^{11}$Li three-body system.

{\it The neutron-neutron interaction} $V_{nn}$ contains central,
spin-orbit, tensor and spin-spin interactions, i.e.
\begin{eqnarray}
V_{nn}(r)= V_c(r) + V_{ss}(r) \bd{s_1} \cdot \bd{s}_2  
 + V_T(r) S_{12}    \nonumber \\ +
        V_{so}(r) \bd{\ell}_{nn} \cdot \bd{s}_{nn} \; ,
\end{eqnarray}
where $\bd{\ell}_{nn}$ is the relative orbital angular momentum,
$S_{12}$ is the tensor operator, $\bd{s}_1$ and $\bd{s}_2$ are the
spins of the two neutrons and $\bd{s}_{nn} = \bd{s}_1 + \bd{s}_2$.  We
assume gaussian shapes for each of the radial potentials $V(r)$. The
parameters are adjusted to reproduce the experimental low-energy
neutron-neutron scattering data for $s$ and $p$-waves.  Different
radial shapes and in general more elaborate potentials have previously
been used in the present context. The results are essentially
indistinguishable from each other provided the scattering lengths and
effective ranges remain unchanged.  We shall therefore use the
interactions specified in \cite{gar97b}, where the corresponding $s$
and $p$-wave scattering lengths and effective ranges also are given.

The {\it neutron-$^{9}$Li interaction} is not very well known,
although several pieces of information are available. The low-energy
properties are almost exclusively determined by $s$ and $p$-waves and
we shall therefore only include these lowest partial waves. The
neutron $p_{3/2}$-state is bound by about 4 MeV in $^{9}$Li and due to
Pauli blocking this state is unavailable for the valence neutron. We
are then left with the $s_{1/2}$ and $p_{1/2}$-states. The total
neutron angular momentum of 1/2 for each of these states then couple
to the $^{9}$Li-spin of 3/2 resulting in two pairs of spin-split
states with angular momenta 1 and 2 and parities corresponding to $s$
and $p$-states, respectively.

The interaction in the present three-body calculations are determined
by the low-energy scattering properties, which in turn determine
positions and widths of resonances and virtual states. These positions
are in fact the decisive properties of the interaction and the
parametrization of the force is in itself rather unimportant. We have
therefore basically only four parameters of physical importance,
i.e. the four positions of the $s$ and $p$-states. Furthermore, the
fragmentation results are only sensitive to the statistically averaged
positions of the $s$ and $p$-states, i.e. the spin-splitting, almost
unavoidable for the strong interaction, does not significantly
influence the computed differential cross sections, see the next
section. The relative position of the $s$ and $p$-states determines
the $p^2$-content in the three-body wave function of $^{11}$Li.  The
$s$ and $p$-wave content are roughly believed to be of the same order
as shown in both calculations and interpretations of experimental
fragmentation data \cite{zin95,zin97,zhu94,gar96,gar97,ber98,mau96}.
This amount of $p$-wave admixture strongly influences the
fragmentation cross sections, which in addition also are very
sensitive to the binding energy or equivalently to the radius of
$^{11}$Li.

Thus we have two crucial parameters ($s$ and $p$-state positions) and
two sensitive observables (three-body binding and $p$-wave
content). However, this perfect match is upset by additional
experimental information, especially knowledge about a $p$-resonance in
the neutron-$^{9}$Li system \cite{zin97,boh93,kry93,you94,abr95}. As
we shall see all three experimental constraints can only be reproduced
simultaneously by use of a three-body interaction, which then is
constructed to add the missing binding energy in the three-body
system.

We shall therefore first concentrate on the neutron-$^{9}$Li two-body
system. We assume central, spin-spin and spin-orbit terms in the
neutron-core interaction, i.e.
\begin{eqnarray}
V_{nc}^{(\ell)}(r)=V_c^{(\ell)}(r)
   +V^{(\ell)}_{ss}(r) \bd{s_n} \cdot \bd{s}_c \nonumber \\
   +V^{(\ell)}_{so}(r) \bd{\ell}_{nc} \cdot \bd{s}_{n} \; , \label{e27}
\end{eqnarray}
where $\bd{\ell}_{nc}$ is the relative orbital angular momentum,
$\bd{s}_n$ and $\bd{s}_c$ are the intrinsic spins of the neutron and
the core. As for the neutron-neutron interaction the radial potentials
$V(r) \propto \exp(-r^2/b^2)$ are gaussians, adjusted independently for
each partial wave.

We choose a large inverse spin-orbit strength to make the
$p_{3/2}$-interaction sufficiently repulsive to avoid any contribution
of this partial wave in the three-body wave function. Furthermore we
shall use a shallow $s$-wave potential without bound states. This
automatically excludes the lowest Pauli forbidden neutron-core
$s$-state from the three-body wave function \cite{zhu93,gar97b}.

As an important constraint we shall use the ``strong evidence'' for a
$p$-resonance at $538\pm 62$ keV with a width of $358\pm 23$ keV
\cite{you94}. We then place the lowest $p_{1/2}$-resonance at 0.5 MeV
with a width of 0.4 MeV consistent with the experimental values.  We
choose rather arbitrarily the total spin of this state to be 1. The
range of the gaussian interaction is in this way determined to be $b=
2.55$ fm. A different range produces different widths of the
resonances. We then continue again somewhat arbitrarily by placing the
other $p_{1/2}$-resonance at 0.92 MeV.  To get a potential at the
limit without a three-body potential we determine the statistically
averaged position of the two $s_{1/2}$-states to be at 350 keV such
that the experimental $^{11}$Li binding energy is reproduced.  This
fixes the strength of the $s$-interaction when we use the same range
as for the $p$-interaction. Finally we use the spin-spin interaction
to place the lowest $s_{1/2}$-state of total spin 2 at an energy of
230 keV. This potential, labeled $II$, leads to a 30\% $p$-wave
content in the $^{11}$Li wave function. We could as well have chosen
the lowest $s_{1/2}$-state with total spin 1 at the same energy of 230
keV. The results for such a potential would be indistiguishable
provided that the average energy of the $s_{1/2}$-states is the same
\cite{gar97}.

We could also have chosen a different position for the highest
$p_{1/2}$-state. However, a lower value than 0.92 MeV would place the
two $p_{1/2}$-states rather close. A higher value would require a
lower statistically averaged position of the $s_{1/2}$-states,
provided the $^{11}$Li binding energy remains unchanged, and the
resulting $p$-wave content then becomes unreasonably small (below
30\%) in disagreement with the fragmentation data
\cite{zin95,zin97,ber98}.

We can now increase the $p$-wave content by increasing the
statistically averaged position of the $s_{1/2}$-states while the
parameters for the $p$-waves remain unchanged. The resulting
underbinding of $^{11}$Li must then be compensated by an attractive
three-body force.  In this way we construct potential I with 40\%
$p$-wave content. 
Again we could have chosen a different position for
the highest $p_{1/2}$-state while maintaining a reasonable $p$-wave
content between 35\% and 40\%.\hfill  However, a lower value would 
\begin{table}
\caption{Parameters for various neutron-$^9$Li interactions. The form
is given in Eq.(\ref{e27}) and the radial shapes are all gaussians,
$\exp(-r^2/b^2)$, with ranges $b = 2.55$ fm and strengths denoted by
$V_{c}^{(\ell)}, V_{ss}^{(\ell)}$ and $V_{so}^{(\ell)}$ with $\ell
=0,1$ for $s$ and $p$-waves, respectively. The strength parameter for
the corresponding three-body interaction, $V_3 \exp(-\rho^2/b_3^2)$
with $b_3=2.50$ fm, are also given. The lower part of the table
contains $p$-wave content in \% in the $^{11}$Li wave function and the
four energies of the resonances and virtual states
$E_{s_{1/2}}^{(1)}$, $E_{s_{1/2}}^{(2)}$, $E_{p_{1/2}}^{(1)}$ and
$E_{p_{1/2}}^{(2)}$.  All the energies and strengths are given in MeV,
and the ranges in fm.  All potentials are defined in the text except
III which is from \protect\cite{cob98}.}
\vspace{0.3cm}
\begin{tabular}{c|ccccc}
   & $I$ & $II$ & $III$ & $IV$ & $V$ \\
\hline
$V_c^{(0)}$ & $-5.60$ & $-6.42$ & $-7.28$ &$-5.60$ & $-5.60$  \\
$V_{ss}^{(0)}$ & $-1.75$ & $-0.75$ & $-0.31$ & 0.00 & -3.00\\
$V_c^{(1)}$ &  $-5.00$ & $-5.00$ & $18.25$ & $-5.00$ & $-5.00$ \\
$V_{ss}^{(1)}$ & 1.00 & 1.00 & 1.47 & 0.00 & 2.00 \\
$V_{so}^{(1)}$ & 33.60 & 33.60 & 55.00 & 33.60 & 33.60 \\
$V_{3}$ & $-3.75$ & 0.00 & 0.00 & $-3.75$ & $-3.75$ \\
\hline
$p^2$(\%) & 40 & 30 & 20 & 40 & 40 \\
$E_{s_{1/2}}^{(2)}$ & 0.24 & 0.23 & 0.14 & 0.58 & 0.09 \\
$E_{s_{1/2}}^{(1)}$ & 1.49 & 0.62 & 0.25 & 0.58 & 2.37 \\
$E_{p_{1/2}}^{(1)}$ & 0.50 & 0.50 & 0.75 & 0.75 & 0.27 \\
$E_{p_{1/2}}^{(2)}$ & 0.92 & 0.92 & 1.60 & 0.75 & 1.13 \\
\end{tabular}
\label{tab1}
\end{table}\noindent
then
require a lower statistically averaged position of the
$s_{1/2}$-states and a repulsive three-body force in order to get the
correct $^{11}$Li binding energy. 
A higher value would 
require a
rather high statistically averaged position of the $s_{1/2}$-states,
i.e. close or even above the statistical average of the
$p_{1/2}$-states. The freedom in choosing the positions of the
resonances and virtual states is in fact limited.

The parameters of the potentials I and II are given in Table
\ref{tab1} along with the $p$-wave content and the energies of the
resonances and virtual states. The statistically averaged position of
the two virtual $s$-states in $^{10}$Li is higher for potential I than
for potential II resulting in the correspondingly larger $p$-wave
content.

In Table \ref{tab1} we also for comparison give the parameters
corresponding to the neutron-$^9$Li interaction (called III) used in
\cite{cob98,cob98a}. The main difference compared to potentials $I$
and $II$ is that the lowest $p$-resonance is placed at 0.75 MeV with a
width of 0.87 MeV.  Maintaining the $p$-interaction and increasing the
$p$-wave content to about 40\% would require a statistically averaged
position of the $s$-states above the corresponding average position of
the $p$-states. This is the reason for the choice of the slightly
lower value of 0.5 MeV for the lowest $p$-state. Potential I is a
rather good starting point. To investigate the effects of
spin-splitting we also constructed the potentials IV and V, where the
statistically averaged positions of the $s$ and $p$-waves are
maintained. Then the $p$-wave content and the $^{11}$Li binding energy
is also unchanged.

In conclusion, the neutron-$^{9}$Li potential is already rather
severely constraint by existing data. The lowest $p_{1/2}$-state is
experimentally determined to be around 0.5 MeV. The ground state is an
$s$-state.  The distance between the statistically averaged positions
for the $s_{1/2}$ and $p_{1/2}$-states is determined by the
requirement of about 40\% $p$-wave content of the neutron-core
relative motion in the total $^{11}$Li wave function. For example
maintaining the $^{11}$Li binding energy at around 300 keV, the
$p$-wave content around 40\% and the three-body force in potential I, a
$p_{1/2}$ average energy at around 0.65 MeV would require the
$s_{1/2}$ average energy at roughly the same value.  If the lowest
$p_{1/2}$-state is at 0.5 MeV the highest $p_{1/2}$-state has to be
above 0.72 MeV, because the $p_{1/2}$ average energy otherwise would
fall below the $s_{1/2}$ average energy. In the same way, a low
$s_{1/2}$-state close to zero requires the highest $s_{1/2}$-state
below 1.7 MeV to avoid the same inversion.

The {\it three-body interaction} is chosen as $V_3
\exp(-\rho^2/b_3^2)$, where $b_3 = 2.50$ fm and the strength $V_3$ is
adjusted to reproduce the measured two-neutron separation energy for
$^{11}$Li, i.e. 295 $\pm$ 35 keV. This additional force is necessary,
since the two-body interactions reproducing all low-energy scattering
phase shifts lead to an underbinding of the three-body system. For
$^{6}$He this deficiency amounts to around 0.5 MeV
\cite{cob98a,cob97}. The idea is that the three-body force should
account for the polarization of the particles beyond that described by
the two-body interactions. Then all three particles must be close to
produce this additional polarization or modification of the intrinsic
structure of the composite particles.  Thus the three-body force must
be of short range.

The importance of the three-body interaction is perhaps seen most
clearly in three-body continuum calculations where the resonance
structure of the two-body subsystems probably is decisive. The
three-body force is designed to give the correct three-body binding
energy while the two-body interactions remain unchanged still
reproducing the two-body structure. With the correct two-body
interactions the computed three-body continuum structure is much more
reliable \cite{fed97}.

For the {\it neutron-carbon interaction} we use non-relativistic
optical potentials obtained from relativistic potentials through a
reduction of the Dirac equation into a Schr\"{o}dinger-like equation
\cite{udi95}. These phenomenological potentials in the Schr\"{o}dinger
equation produce the same scattering data as obtained by use of the
relativistic potentials in the Dirac equation \cite{she86}.

\begin{figure}[ht]
\centerline{\psfig{figure=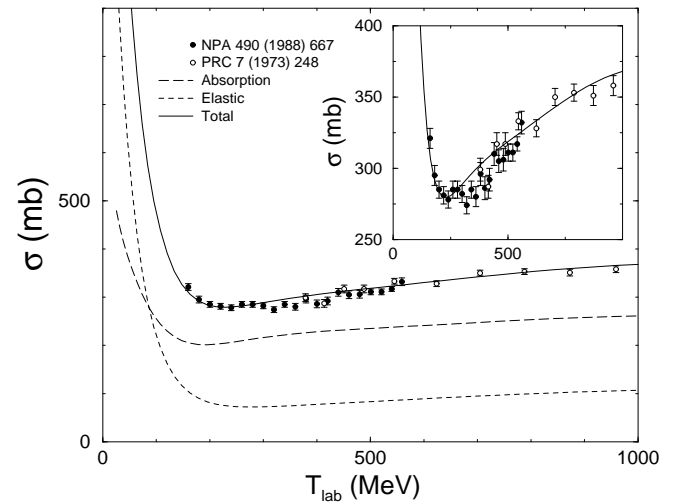,width=8.5cm,%
bbllx=2.7cm,bblly=3cm,bburx=20cm,bbury=25.3cm,angle=270}}
\vspace{0.2cm}
\caption[]{The total (solid), elastic (dashed) and absorption
(short-dashed) cross sections $\sigma$ for neutron-$^{12}$C reactions
as functions of the neutron laboratory kinetic energy $T_{lab}$. The
inset focuses on the region of the minimum.  The open and filled
circles are experimental points from \protect\cite{dev73,fra88},
respectively. The curves are obtained in an optical model computation
as decribed in \protect\cite{udi95} using the parameters in
\protect\cite{coo93}. We include 35 partial waves in the
calculations.}
\label{fig2}
\end{figure}

In particular for a carbon target the interaction used is the
parameterization EDAI-C12 \cite{coo93} valid for a range of neutron
kinetic energies from 29 to 1040 MeV.  The resulting neutron-$^{12}$C
cross sections are shown in Fig. \ref{fig2}. For energies above 150
MeV the cross sections are relatively weak functions of energy and
absorption is much more likely than diffraction. The factorization of
the fragmentation cross sections in Eqs.(\ref{eq5}) and (\ref{eq6}) is
then rather accurate at these high energies.

At energies below 150 MeV the neutron-target cross sections increase
dramatically.  The factorization is then less accurate, but still a
useful approximation.  Furthermore, diffraction quickly dominates over
absorption. The absolute cross sections for fragmentation of $^{11}$Li
can then be expected to increase with decreasing beam energy. The
shapes of the distributions also should change from narrow
(absorption) to broad (elastic scattering) with decreasing energy. At
intermediate energies we can expect a mixture exhibiting a narrow
distribution with a relatively large broad background. These
predictions are simple consequences of the model, where the
fragmentation cross sections essentially are proportional to the
neutron-carbon cross sections.

The {\it carbon-$^{9}$Li interaction} is needed to estimate the
cross sections both when the $^{9}$Li-core is destroyed
(absorped) and the two neutrons detected and when the core is
scattered (diffracted) on the target. These cross sections are
obtained by a two-body computation of the carbon-$^{9}$Li reactions
using the simple almost schematic optical potential defined in
\cite{kar75} and described in \cite{ber93}.  In this model the beam
energy dependence is introduced through the proton-proton and
neutron-proton cross sections. Experimental data of these
nucleon-nucleon cross sections can be found in \cite{par73}.

In the calculation of the $^9$Li-$^{12}$C cross section we assume spin
zero for $^9$Li. We need around 150 partial waves to get convergence.
At a beam energy of 280 MeV/u the computed elastic cross section is
419 mb and the computed absorption cross section is 795 mb. This gives
a total of 1214 mb consistent with the results in \cite{pen81}. These
values are dramatically reduced by the shadowing effect.

\section{Comparison with Experiments}

The interaction parameters discussed in the previous section determine
the structure of the three-body projectile. The positions of the
resonances and the $p$-wave content are the crucial quantities. The
reactions are described by the participant-target interactions and the
cutoff radii taking the shadowing effect into account. We previously
used the cutoff radii $r_{nc}=r_{nn}=3$ fm \cite{gar98b}, which is
consistent with the fragmentation data for $^6$He on carbon
\cite{gar98}. For $^{11}$Li fragmentation the neutron-neutron
shadowing parameter should remain the same, since the physical origin
is unchanged. Therefore we use $r_{nn}=3$ fm while we still maintain
the neutron-$^9$Li shadowing radius $r_{nc}$ as a parameter for
adjustments within rather narrow limits around the value obtained from
Eq.(\ref{e29}). The radius of $^9$Li is 0.9 fm larger than the radius
of $^4$He and the value of $r_{nc}$ is then expected to be about 4 fm
for $^{11}$Li. In this section we shall compute different types of
observables as discussed in the previous section and use the
experimental data to select a promising set of interactions and
shadowing parameters. We shall only display results for potentials I,
II and III and omit the curves for potentials IV and V, which in all
cases barely can be distinguished from those of I.

The {\it two-neutron removal cross section} $\sigma_{-2n}$ is known
experimentally for $^{11}$Li fragmentation on a carbon target at 280
MeV/u \cite{zin97}.  All events, where $^9$Li is detected after the
fragmentation, contribute to this cross section.  According to
\cite{zin97} three different reaction mechanisms lead to such a halo
breakup. First electromagnetic excitation of the halo state into the
continuum followed by decay of these excited states. We shall neglect
this process, since it is expected to give a rather small contribution
(less than 10 mb) for a light target like carbon.

The second mechanism is the stripping of one of the halo neutrons by
the target. This process is also called absorption and the
corresponding cross section is denoted by $\sigma_{-2n}^S$. It is
fully described in our model via the imaginary part of the
neutron-target optical potential. On the other hand we neglect
processes where both neutrons are absorbed by the target and no
neutrons appear in the final state. This contribution to the cross
section is expected to be small, since the large spatial extension of
the projectile diminishes these simultaneous reactions.

\begin{table}
\hspace*{0.8cm}
\caption{Core diffraction ($\sigma_{-2n}^{CD}$), two-neutron removal
cross sections ($\sigma_{-2n}$) equal to the sum of neutron
diffraction ($\sigma_{-2n}^D$) and neutron stripping
($\sigma_{-2n}^S$) computed for fragmentation of $^{11}$Li at an
energy of 280 MeV/u. The lowest part of the table is for an aluminum
target whereas everything else refers to a carbon target. The cross
sections are in millibarns. The potentials and shadowing parameters
are from Table \ref{tab1}. The experimental data are from
\protect\cite{zin97} and \protect\cite{hum95} for carbon and aluminum,
respectively. Potentials IV and V produce the same values as potential
I.}
\vspace{0.3cm}
\begin{tabular}{ccc|cccc}
Int & $r_{nc}$ & $r_{nn}$ & $\sigma_{-2n}^{CD}$ & $\sigma_{-2n}^D$ & 
$\sigma_{-2n}^S$ & $\sigma_{-2n}$ \\
\hline
 & C & & & $80\pm20$ & $200\pm20$ & $280\pm30$ \\
\hline
 I  & 0 & 0 & 88 & 146 & 437 & 583 \\
 II & 0 & 0 & 88 & 146 & 437 & 583 \\
III & 0 & 0 & 89 & 145 & 436 & 581 \\
\hline
 I  & 3.5 & 3 & 63 & 68 & 204 & 272 \\
 II & 3.5 & 3 & 65 & 72 & 215 & 287 \\
III & 3.5 & 3 & 65 & 76 & 227 & 303 \\
\hline
 I  & 4 & 3 & 52 & 61 & 184 & 245 \\
 II & 4 & 3 & 55 & 65 & 195 & 260 \\
III & 4 & 3 & 55 & 69 & 207 & 276 \\
\hline
 & Al & & & - & - & $470 \pm 80$ \\
\hline
  I & 0 & 0 & 107 & 358 & 901 & 1259 \\
  I & 4 & 3 &  70 & 150 & 379 &  529 \\
  I & 5 & 4 &  52 & 100 & 251 &  351 \\
  I & 6 & 5 &  38 &  68 & 172 &  240 \\
\end{tabular}
\label{tab2}
\end{table}

The third mechanism is the diffraction of one of the halo neutrons. We
also refer to this process as elastic scattering and we denote the
corresponding cross section by $\sigma_{-2n}^D$.  For weakly bound
systems and not too small beam energy the main contribution to
$\sigma_{-2n}^D$ comes from processes where the neutron after
scattering on the target ceases to interact with the remaining two
halo particles. This is precisely our model, where the final state is
described as two independent two-body systems and the interaction
between them is ignored, see Fig. \ref{fig1}.

In addition to these three mechanisms there must be a contribution
($\sigma_{-2n}^{CD}$) where the core is scattered by the target. In
Table \ref{tab2} we give these computed cross sections for the
potentials in Table \ref{tab1} and different sets of shadowing
parameters. It is remarkable that the cross sections calculated
without shadowing are virtually independent of the potentials, and
furthermore clearly much larger than observed. The computed values are
reduced by roughly a factor of two by using the shadowing parameters
$r_{nc}=3.5$ fm and $r_{nn}=3$ fm. Then by neglecting the two-neutron
absorption (not computed) and the core diffraction processes the
experimental numbers are reproduced within the error bars for the
total as well as for the stripping and diffraction cross sections.
Therefore we should underestimate the experimental values and use the
larger shadowing parameters $r_{nc}=4$ fm and $r_{nn}=3$ consistent
with the parametrization in Eq.(\ref{e29}).

\begin{figure}[ht]
\centerline{\psfig{figure=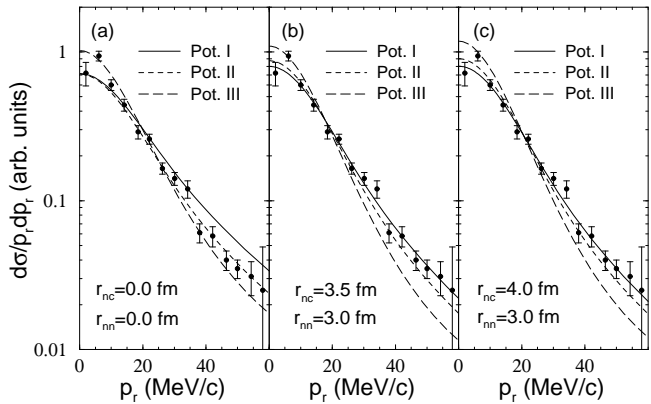,width=8.5cm,%
bbllx=3.7cm,bblly=2.1cm,bburx=17.8cm,bbury=24.6cm,angle=270}}
\vspace{0.2cm}
\caption[]{Radial neutron momentum distribution after fragmentation of
$^{11}$Li on a carbon target at 280 MeV/u. Core diffraction is not
included. The experimental data from \protect\cite{zin95} is compared
with calculations using the potentials I (solid), II (short-dashed)
and III (long-dashed) in Table \ref{tab1}. The shadowing parameters
are given on the figure.  The optical model parameters are from
\protect\cite{coo93}.}
\label{fig3}
\end{figure}

However, the computed core diffraction contributions in Table
\ref{tab2} are not negligible, but for comparison with experimental
values they should be reduced, since (i) a fraction corresponds to
elastically scattered $^{11}$Li-particles, (ii) the large transverse
momenta of $^{9}$Li precisely obtained through the scattering process
to a large extent are experimentally excluded due to limitations in
the large angle acceptance, (iii) the shadowing by the core is
probably larger than the shadowing by a neutron corresponding perhaps
rather to $r_{nc}=5$ fm and $r_{nn}=4$. Furthermore, the calculations
rely on rather uncertain optical model parameters. The effect is
almost entirely confined to the absolute values while the shapes of
the distributions only are marginally influenced. Using the complete
factorization approximation we estimate the size of this cross section
to be about 60\% of the values of $\sigma_{-2n}^{CD}$ in Table
\ref{tab2}. If necessary it can be added at the appropriate
places, but we shall in the remaining part of this paper not include
this process unless explicitly mentioned.

The radial {\it neutron momentum distribution after neutron removal}
in fragmentation of a 280 MeV/u $^{11}$Li-projectile on carbon is
measured \cite{zin95}. The variable is $p_r$, where $p_r^2= p_x^2 +
p_y^2$ is expressed in terms of the projections $p_x$ and $p_y$ of the
neutron momentum along the two directions perpendicular to the beam
direction chosen as the $z$-axis. In Fig. \ref{fig3} we compare our
calculations (suitably scaled) for different potentials with the
measured momentum distributions in arbitrary units. Without shadowing,
shown in part (a), potentials $I$ and $III$ give too broad and too
narrow distributions, respectively, whereas potential $II$ with 30\%
p-wave content reproduces the data.  This result is consistent with
previous computations without shadowing and diffraction
\cite{gar96,gar97}.  With shadowing, shown in parts (b) and (c), we
obtain as expected narrower distributions. Variation from $r_{nc}=3.5$
fm to 4 fm is hardly visible showing a very weak dependence on the
$r_{nc}$ shadowing parameter. Now potential $III$ gives a too narrow
neutron momentum distributions while potentials $I$ and $II$ both
reproduce the data.

\begin{figure}[ht]
\centerline{\psfig{figure=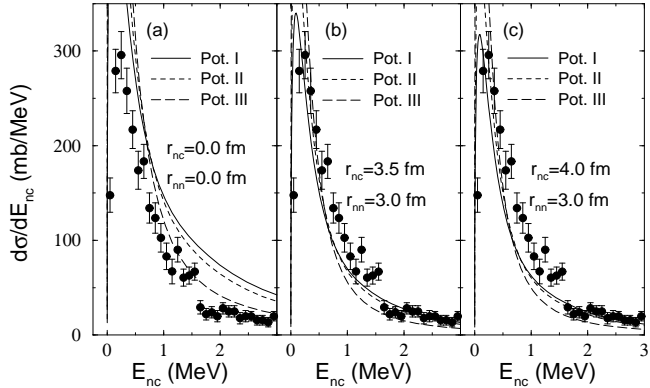,width=8.5cm,%
bbllx=3.7cm,bblly=1.5cm,bburx=17.8cm,bbury=24.7cm,angle=270}}
\vspace{0.2cm}
\caption[]{Invariant mass spectrum of $^{10}$Li after fragmentation of
$^{11}$Li on a carbon target at 280 MeV/u. Only neutron and core
spectators forming $^{10}$Li are included. The experimental data are
from \protect\cite{zin97}. The calculated curves in (a), (b) and (c)
are obtained with the same parameters as in Fig. \ref{fig2}. The
maxima for the computed curves in panel (a) (outside the figure) are
around 500 mb/MeV, 650 mb/MeV and 1000 mb/MeV for potentials $I$, $II$
and $III$, respectively. In panel (b) and (c) the maxima for potential
$II$ are 430 mb/MeV and 400 mb/MeV, respectively. For potential $III$
we obtain instead 650 mb/MeV and 600 mb/MeV. }
\label{fig4}
\end{figure}

The {\it invariant mass spectrum} of $^{10}$Li is independent of the
coordinate system and in the rest system of $^{10}$Li the invariant
mass is (after subtraction of the rest masses) equal to the total
kinetic energy $E_{nc}$ of the neutron-$^{9}$Li system
\cite{zin97,gar97b}. In Fig. \ref{fig4} we compare calculations with
experimental data obtained after fragmentation of $^{11}$Li
\cite{zin97}. The experimental distribution is given in absolute
numbers and we compare directly without any arbitrary scaling. We
only include processes where the spectators form the detected
$^{10}$Li, i.e. we neglect those neutron-core combinations where the
neutron or the core have been through the scattering process. The
latter contributions are relatively small, i.e. the core diffraction
and half of the neutron diffraction cross sections. Furthermore the
corresponding contributions from these participants are probably not
fully included in the measurement.

The calculations without shadowing in panel (a) produce for all
potentials distributions with maxima much higher than observed.  With
shadowing the maxima are reduced as shown in panels (b) and (c), which
again are rather similar, but still potentials $II$ and $III$ both
give too high peaks. Clearly the best comparison is obtained for
potential $I$ and especially with the shadowing in panel (c). The
discrepancies appear as a slightly too low-lying peak energy and a
slightly too narrow peak. However, the response function of the
detector system is not accounted for and a more accurate comparison
would shift the peak towards higher values, decrease the maximum value
and broaden the peak \cite{zin97}. Thus all defiencies may be improved
in this way. In any case the behavior of the invariant mass spectrum
is strongly influenced by the properties of the low-lying
resonances. For example a narrow $p$-resonance would produce a
shoulder or a peak in the distribution at the position of the
resonance. A broader resonance would only show up as a larger cross
section at the corresponding position. Thus a more detailed and more
accurate comparison between computed and measured distributions would
provide information about the neutron-core resonance structure.

The radial {\it neutron momentum distribution after core breakup} of
$^{11}$Li on a carbon target at 280 MeV/u is also known experimentally
for collisions where $^{9}$Li is destroyed \cite{nil95}. The neutrons
are detected in coincidence with a charged fragment different from
$^9$Li. The contribution of neutrons coming from the core is
eliminated in the experiment by subtraction of the corresponding
neutron momentum distribution obtained from the measured $^9$Li-carbon
reaction.

This distribution arises only from absorption, i.e. we must compute
the cross section in Eq.(\ref{eq6}), where the $^9$Li-core is
destroyed by the target while the neutrons continue unaffected. For
simplicity we assume complete factorization such that the
$^9$Li-carbon interaction only is necessary to provide the absolute
scale through the absorption cross section. This is a very good
approximation for large beam energies. The momentum distribution is
then given by Eq.(\ref{sudden}) and therefore computed as described in
\cite{gar97}. The absolute cross section is obtained afterwards by
multiplication of the absorption cross section computed with the
optical model parameters in \cite{kar75,ber93} to be 795 mb for a beam
energy of 280 MeV/u.

\begin{figure}[ht]
\centerline{\psfig{figure=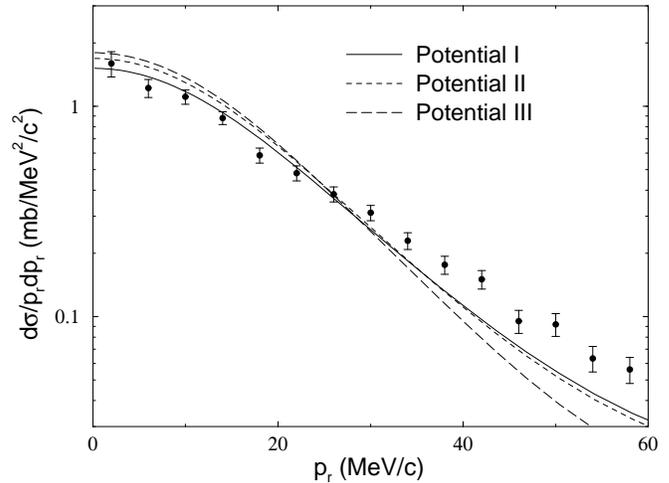,width=8.5cm,%
bbllx=3.7cm,bblly=2.1cm,bburx=19.9cm,bbury=23.6cm,angle=270}}
\vspace{0.2cm}
\caption[]{Radial neutron momentum distribution after core breakup
fragmentation of $^{11}$Li on a carbon target at 280 MeV/u. The
calculations are for the interactions in Table \ref{tab1} and a
neutron-core shadowing parameter of 4 fm. The $^{9}$Li-carbon
interaction providing the absolute values is the optical model from
\cite{kar75,ber93}. The experimental data \protect\cite{nil95} in
arbitrary units are scaled to match the calculations. }
\label{fig5}
\end{figure}

Both from Table \ref{tab2} and Figs. \ref{fig3} and \ref{fig4} we
found good agreement between theory and experiment with the choice
$r_{nc}=4$ fm and $r_{nn}=3$ fm for the shadowing parameters.  For
core breakup reactions, where only the neutron-core shadowing
parameter is relevant, we therefore use $r_{nc}=4$ fm in the
computation of the neutron momentum distribution shown in
Fig. \ref{fig5}. The agreement between theory and experiment is
remarkably good in view of the simplicity of the model, where the
interactions between the halo neutrons and all the fragments from the
core destruction have been neglected. In contrast to the other
observables this reaction produces fragments with approximately the
same velocity as the halo neutrons and final state interactions beside
that of the two neutrons could be significant. The two spectator
neutrons could also be disturbed during such violent reactions.

The larger $p$-wave content arising for potential $I$ gives a broader
distribution closer to the data than the potentials $II$ and $III$.
The comparison in Fig. \ref{fig5} is then also supporting the choice
of potential $I$ and a neutron-core shadowing parameter of 4 fm in the
description of the fragmentation reactions of $^{11}$Li on a light
target.

\begin{figure}[ht]
\centerline{\psfig{figure=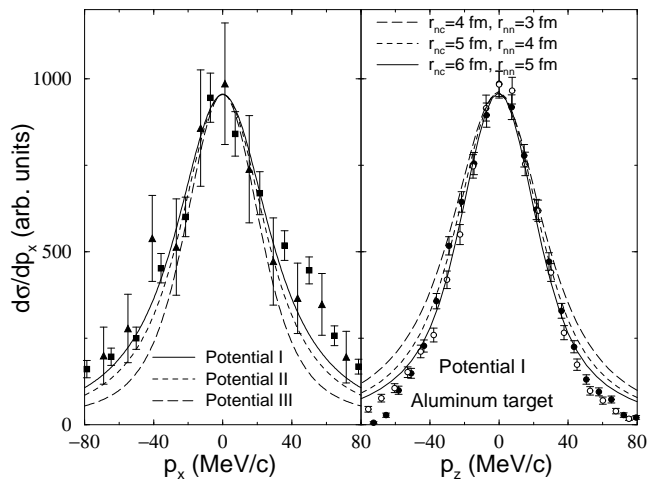,width=8.5cm,%
bbllx=3.7cm,bblly=2.0cm,bburx=19.9cm,bbury=23.6cm,angle=270}}
\vspace{0.2cm}
\caption[]{Momentum distributions for the $^9$Li-core after
fragmentation of $^{11}$Li. Core diffraction is not included. The left
hand side is the transverse distribution for a beam energy of 280
MeV/u on a carbon target and the experimental data is from
\protect\cite{hum95} (filled squares) and \cite{zhu95} (filled
triangles). The calculations are for the potentials in Table
\ref{tab1} with ($r_{nc},r_{nn}$) = (4 fm, 3 fm). The right hand side
shows the experimental longitudinal distribution for a beam energy of
468 MeV/u (filled circles) and 648 MeV/u (open circles) on an aluminum
target \protect\cite{gei97}. The calculations are for an energy of 280
MeV/u for potential $I$ and the three sets of shadowing parameters
specified on the figure. The optical model parameters are from
\protect\cite{coo93}. All the computed distributions are scaled to the
experiment. }
\label{fig6}
\end{figure}

The transverse $^9$Li-{\it core momentum distribution} for breakup of
$^{11}$Li on a carbon target at 280 MeV/u is computed and shown in
Fig. \ref{fig6} for the potentials and shadowing parameters in Table
\ref{tab1}. Detailed comparison with the data \cite{hum95,gei97}
requires folding of the computed curves with the experimental beam
profile resulting in a few MeV broader curves \cite{zhu95}. Due to the
large experimental errors this distribution is not very helpful in
constraining the potentials and shadowing parameters. However, we can
still conclude that the agreement between theory and experiment is
satisfactory for potential $I$.

In Fig. \ref{fig6} we also show the experimental longitudinal core
momentum distribution on an aluminum target for two different beam
energies.  In the next section we shall show that the difference
between longitudinal and transverse momentum distributions is rather
small and the widths of the distributions are almost independent of
the beam energy. The difference between the transverse and
longitudinal data in Fig. \ref{fig6} can be explained by the use of a
different target in the experiments. The aluminum radius is almost one
fermi larger than the carbon radius, the optical model parameters are
different and more important the shadowing parameters must be larger
for aluminum with the resulting narrower momentum distributions as
seen in Fig. \ref{fig6}. Still the tail of the distribution is not
reproduced. The theoretical prediction of the two-neutron removal
cross sections for fragmentation of $^{11}$Li on an aluminum target
are shown in Table \ref{tab2}. The parameters $(r_{nc},r_{nn})$ = (5
fm, 4 fm) are consistent with Eq.(\ref{e29}) and the corresponding
two-neutron removal cross section is then expected to be $\sigma_{-2n}
\approx 350$ mb.

The conclusion of this section is that potential I with the shadowing
parameters $(r_{nc},r_{nn})$ = (4 fm, 3 fm) for a carbon target gives
an excellent agreement between theory and experiment for the
observables discussed. This is a strong justification for the model
and the method.

\section{Beam Energy Dependence}
We have now established a model with a set of parameters successfully
reproducing a variety of experimental data, i.e. potential I,
$r_{nc}=4$ fm, $r_{nn}=3$ fm and optical model parameters from
\cite{coo93,ber93}. In the remaining part of this paper we shall only
use these parameters and explore the consequences of the model for a
number of observables. Particularly we shall in the following
concentrate on the predicted energy dependence of various quantities
in fragmentation reactions of $^{11}$Li on carbon.  The model only has
a dependence on the beam energy through the interaction between the
participant and the target. This interaction is described by
phenomenological optical models, which give absorption and elastic
scattering cross sections as functions of particle energy, see
Fig. \ref{fig2}. These cross sections are decisive factors in
Eqs.(\ref{eq5}) and (\ref{eq6}) and two-neutron removal cross sections
must show the same energy dependence.

\begin{figure}[ht]
\centerline{\psfig{figure=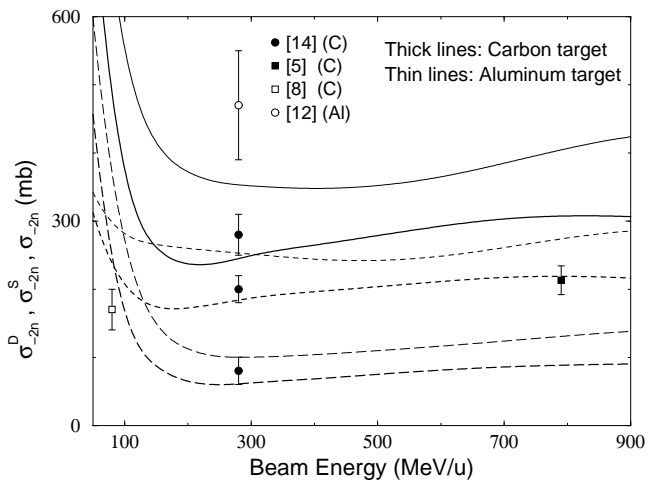,width=8.5cm,%
bbllx=3.7cm,bblly=1.6cm,bburx=19.7cm,bbury=23.8cm,angle=270}}
\vspace{0.2cm}
\caption[]{Two-neutron removal cross sections for $^{11}$Li
fragmentation on carbon (thick curves) and aluminum (thin curves) as
functions of the beam energy. The cross sections from neutron
stripping $\sigma_{-2n}^S$ (short-dashed) and neutron diffraction
$\sigma_{-2n}^D$ (long-dashed) are shown along with the sum
$\sigma_{-2n}$ (solid).  The shadowing parameters are
$(r_{nc},r_{nn})=(4,3)$ fm for carbon and $(5,4)$ fm for aluminum, the
interactions are potential I from Table \ref{tab1} and the optical
model is specified in \protect\cite{coo93,ber93}. The experimental
data are from \protect\cite{bla93,zin97,kob88} for increasing energies
for carbon and from \protect\cite{hum95} for aluminum. For
\protect\cite{kob88,bla93} only the total cross section is given.  }
\label{fig7}
\end{figure}

This is indeed seen in Fig. \ref{fig7} where we show two-neutron
removal cross sections as functions of the beam energy. These
calculations do not include two-neutron absorption processes and
processes where the core interacts with the target. The cross sections
are clearly governed by the behavior of the neutron-carbon cross
section, with a minimum at a beam energy of around 250 MeV/u. For
larger energies we observe smooth increases towards a flat region. For
smaller energies the cross sections increase rather dramatically. The
prediction is an increase by about 70\% when carbon and aluminum are
interchanged as target. The computed curves underestimate the latest
experimental points \cite{zin97}, as expected due to the neglect of
core diffraction. On the other hand the older data are far below the
calculations. However, at 30 MeV/u total two-neutron removal cross
sections are measured for targets of beryllium $0.47 \pm 0.10$ b and
nickel $1.3 \pm 0.4$ b \cite{rii92}. This is in agreement with the
computed increase towards smaller beam energies.

The two-neutron removal cross sections can be separated into a number
of differential cross sections. We shall discuss the momentum
distributions of the halo particles both relative to each other and
individually with respect to the center of mass of the projectile. In
addition we shall discuss the invariant mass of $^{11}$Li and the
angular correlation of the emitted neutrons. These observables are in
most cases not experimentally available and our results are therefore
model predictions.

\begin{figure}[ht]
\centerline{\psfig{figure=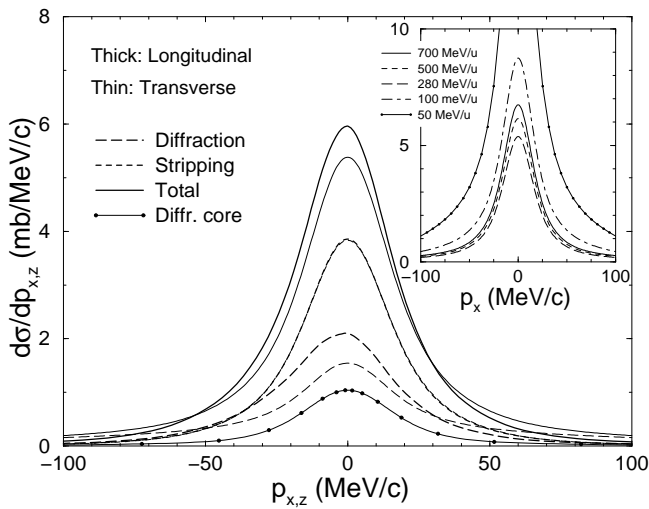,width=8.5cm,%
bbllx=2.5cm,bblly=2.8cm,bburx=19.0cm,bbury=23.8cm,angle=270}}
\vspace{0.2cm}
\caption[]{Longitudinal (thick) and transverse (thin) neutron momentum
distributions in coincidence with $^{9}$Li after fragmentation of
$^{11}$Li on carbon at 280 MeV/u computed in the rest frame of the
projectile. The beam momentum is the reference direction. The
short-dashed and long-dashed curves are the contributions to the total
(solid) from diffraction and stripping of the participant
neutron. Core diffraction is shown separately as the filled
circles. The inset shows the total transverse neutron momentum
distribution for the same reaction for different beam energies in the
same unit. The standard set of parameters in Fig. \ref{fig7} is
used. }
\label{fig8}
\end{figure}

In Fig. \ref{fig8} we show the neutron momentum distributions after
fragmentation of $^{11}$Li on carbon. When the participant neutron is
absorbed in the stripping process, the detected neutron must be a
spectator and then there is no difference between longitudinal and
transverse momentum distributions in the rest frame of the
projectile. When the participant neutron is scattered by the target
(diffraction) it receives additional momentum perpendicular to the
direction of the beam.  This process therefore contributes with a
broader momentum distribution in the transverse than in the
longitudinal direction. The tail in the total neutron momentum
distribution is then more pronounced for the transverse than for the
longitudinal distribution. On the other hand different tails do not
necessarily imply that the full width at half maximum (FWHM) of the
distributions also differs. The relative size of the stripping and
diffraction contributions reflects the size of the corresponding cross
sections in Fig. \ref{fig7}.  The contribution to the neutron momentum
distribution from the core diffraction (dotted line in the figure) is
the same in the longitudinal and transverse directions, as expected
because both neutrons are spectators. This contribution of about 30 mb
is not added in the figure to the total, but the total width would
only be marginally influenced.

The inset in Fig. \ref{fig8} shows the transverse neutron momentum
distribution for different beam energies. The shape of the
distributions is essentially independent of the energy due to the
approximate factorization in Eqs.(\ref{eq5}) and (\ref{eq6}) at these
fairly high energies, see Table \ref{tab3}. The computed FWHM
decreases slightly with an energy increase from 50 MeV/u to 280 MeV/u
and remains then essentially constant at higher energies. In contrast,
the maximum or peak value for the momentum distributions changes
considerably with the beam energy. The behavior of the peak values,
i.e. a sharp decrease, the passing of a minimum around 250 MeV/u
followed by a smooth increase is as expected similar to the variation
shown in Fig. \ref{fig2}.

\begin{table}
\hspace*{0.8cm}
\caption{The maximum values in mb/MeV/c (one-dimensional) and
mb/MeV$^2$/c$^2$ (radial) and the FWHM in MeV/c as functions of beam
energy in MeV/u for various momentum distributions computed for
fragmentation of $^{11}$Li on carbon in the rest frame of the
projectile. Core diffraction is not included. The indices $\bot$, $\|$
and $r$ indicate transverse, longitudinal and radial distributions,
respectively. For the radial distributions the FWHM is the width at
half maximum multiplied by two. For the invariant mass spectrum we
give the maximum value in mb/MeV and the width in MeV.}
\vspace{0.3cm}
\begin{tabular}{c|c|ccccc}
 & Beam energy & 50 & 100 & 280 & 500 & 700 \\
\hline 
 n$_{\bot}$ & max. & 18.4 & 8.7 & 5.4 & 6.2 & 6.7 \\
 n$_{\bot}$ & FWHM & 42 & 41 & 39 & 39 & 39 \\
\hline
 n$_{\|}$ & max. & 21.6 & 10.2 & 6.0 & 7.1 & 7.9 \\
 n$_{\|}$ & FWHM & 42 & 41 & 39 & 39 & 39 \\
\hline
 (n-$^{9}$Li)$_{\bot}=$ & max. & 16.2 & 8.0 & 5.2 & 5.9 & 6.4 \\
 (n-$^{9}$Li)$_{\|}$  & FWHM & 37 & 37 & 37 & 37 & 37 \\
\hline
 n$_{r} \approx $  &  max. & 2.7 & 1.3 & 0.9 & 1.0 & 1.1 \\
 (n-$^{9}$Li)$_{r}$ & FWHM & 31 & 31 & 31 & 31 & 31 \\
\hline
 ($^{9}$Li)$_{\bot}=$ & max. & 9.4 & 4.5 & 2.9 & 3.4 & 3.7  \\
 ($^{9}$Li)$_{\|}$ & FWHM & 63 & 63 & 63 & 63 & 63 \\
\hline
 ($^{10}$Li)$_{\bot}=$ & max. & 12.3 & 6.0 & 3.9 & 4.5 & 4.9 \\
 ($^{10}$Li)$_{\|}$ & FWHM & 49 & 49 & 49 & 49 & 49 \\
\hline
 inv. & max. & 998 & 488 & 317 & 361 & 394 \\
  mass & FWHM & 0.42 & 0.42 & 0.42 & 0.42 & 0.42 \\
\end{tabular}
\label{tab3}
\end{table}

\begin{figure}[ht]
\centerline{\psfig{figure=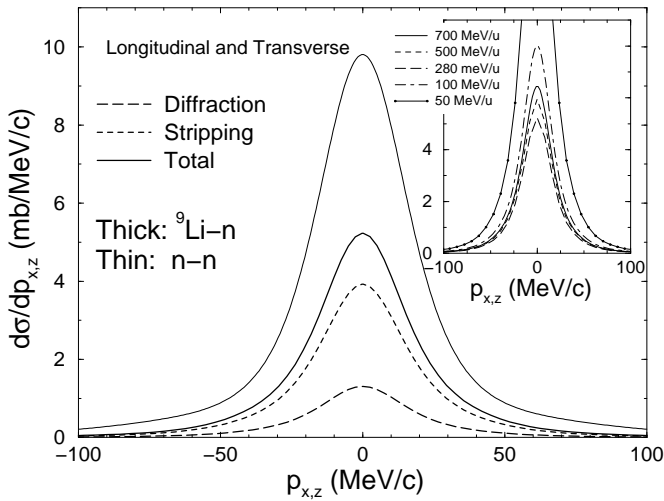,width=8.5cm,%
bbllx=2.5cm,bblly=2.8cm,bburx=19.0cm,bbury=23.8cm,angle=270}}
\vspace{0.2cm}
\caption[]{The neutron-$^9$Li relative momentum distribution for the
cases in Fig. \ref{fig8}. The relatively small contributions from the
neutron and core participants are not included and the longitudinal and
transverse momentum distributions are therefore identical. The
neutron-neutron relative momentum distribution computed for core
destruction using the complete factorization approximation is also
shown for an energy of 280 MeV/u. The carbon-$^9$Li absorption and
diffraction cross sections are 795 mb and 419 mb, respectively. }
\label{fig9}
\end{figure}

Instead of referring the neutron momentum distribution to the rest
system of the projectile as in Fig. \ref{fig8}, we could refer it to
the rest system of the spectator neutron-$^9$Li system. Actually in
this frame the neutron momentum is the relative neutron-$^9$Li
momentum. This momentum distribution is shown in Fig. \ref{fig9}, and
it is identical to those of Fig. \ref{fig8} for an infinitely heavy
core. In Fig. \ref{fig9} we only include the dominating term arising
from the spectator neutron and the longitudinal and transverse
distributions are therefore identical. The neglected contribution is
about half of the diffraction part. The FWHM and the peak values for
the curves in the inset are given in Table \ref{tab3}. The FWHM is 37
MeV/c for all the energies coinciding with the energy independent
width of the stripping part in Fig. \ref{fig8}.

For comparison the neutron-neutron relative momentum distribution
computed using the complete factorization approximation is also shown
in Fig. \ref{fig9}. The shape is in this approximation energy
independent and the scale is determined by the carbon-$^{9}$Li
absorption (core destruction) cross section. For diffraction (core
survival) we should multiply it by about a factor 0.53 for a beam energy
of 280 MeV/u.

\begin{figure}[ht]
\centerline{\psfig{figure=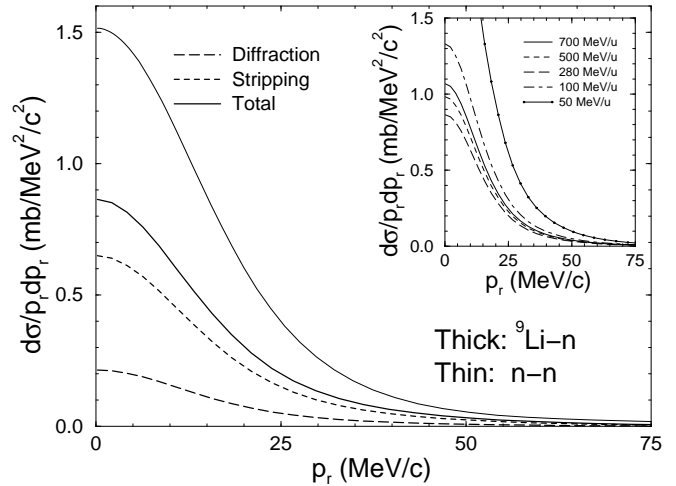,width=8.5cm,%
bbllx=2.8cm,bblly=2.0cm,bburx=18.8cm,bbury=23.5cm,angle=270}}
\vspace{0.2cm}
\caption[]{The radial distributions of the spectator neutron-$^9$Li
and the neutron-neutron relative momenta for the cases in
Fig. \ref{fig9}. }
\label{fig10}
\end{figure}

The two-dimensional radial momentum distributions are often used to
increase the number of observed events. The variable is then $p_r$
($p_r^2=p_x^2+p_y^2$) and integration of this momentum distribution
over $p_x$ (or $p_y$) gives the transverse momentum. The results are
shown in Fig. \ref{fig10} for the cases in Fig. \ref{fig9}. The
widths and the peak values for the curves in the inset are given in
Table \ref{tab3}. We find the same qualitative behavior as for the
one-dimensional distributions. This variation as well as the relative
size of the stripping and diffraction contributions are again
consistent with the result in Fig. \ref{fig2}.

The radial neutron momentum distribution analogous to Fig. \ref{fig8}
is almost indistinguishable from the results in Fig. \ref{fig10}. The
cross section for the neutron-neutron momentum distribution
corresponds to core destruction for a beam energy of 280 MeV/u. The
core survival process is obtained by multiplication with the factor
0.53. The widths are energy independent in this approximation.

\begin{figure}[ht]
\centerline{\psfig{figure=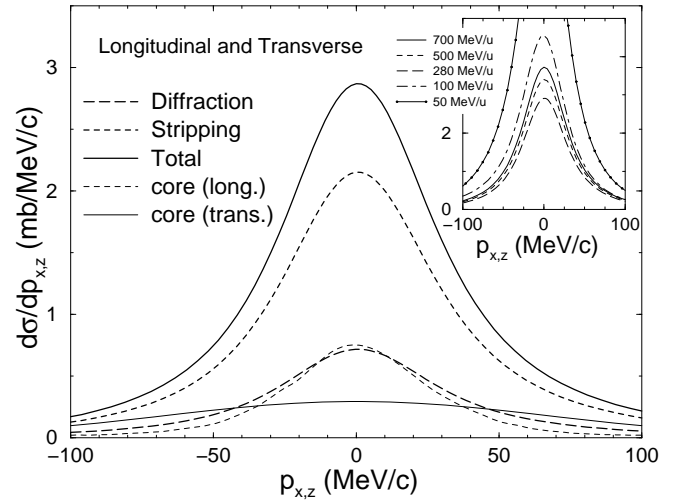,width=8.5cm,%
bbllx=2.8cm,bblly=2.9cm,bburx=19.0cm,bbury=23.8cm,angle=270}}
\vspace{0.2cm}
\caption[]{The $^9$Li momentum distribution for the cases in
Fig. \ref{fig8}. The transverse and longitudinal contributions from
the diffracted core are shown separately, but not included in the total
cross section. The longitudinal and transverse momentum distributions
are therefore identical. The inset shows the distributions in the same
unit for different beam energies.}
\label{fig11}
\end{figure}

The neutron momentum distribution is narrower than that of the $^9$Li
core due to the final state interaction. We show in Fig. \ref{fig11}
the computed $^9$Li momentum distributions for the same cases as in
Fig. \ref{fig8}. The transverse distribution from the diffracted core
is small and very broad due to the diffraction process. The
longitudinal distribution is as usual narrower. These contributions
add about 30 mb to the total cross section while changing only
marginally the shape of the total distribution. They are not added in
the figure where the total distribution then only includes
contributions from the participant neutrons. The displayed
longitudinal and transverse momentum distributions are therefore
identical, since the difference between them is due to the diffraction
process. The inset shows the core momentum distributions for different
beam energies. The computed widths and the peak values are given in
Table \ref{tab3}. The behavior is again a reflection of the results in
Fig. \ref{fig2}.

\begin{figure}[ht]
\centerline{\psfig{figure=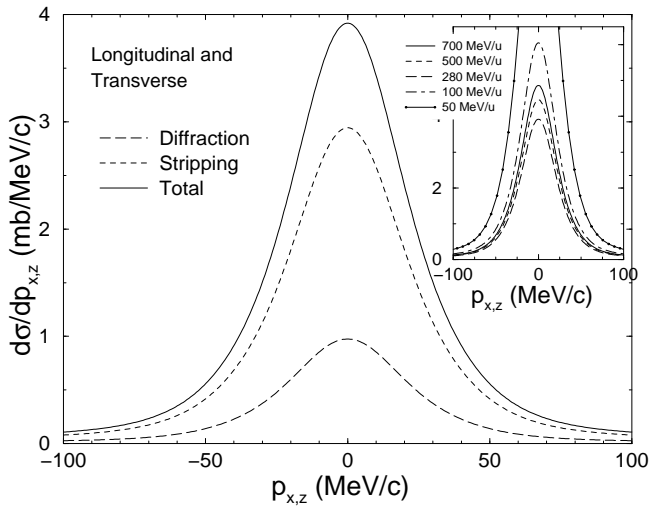,width=8.5cm,%
bbllx=2.5cm,bblly=2.8cm,bburx=18.8cm,bbury=23.8cm,angle=270}}
\vspace{0.2cm}
\caption[]{The momentum distribution for the $^{10}$Li center of mass
for the cases in Fig. \ref{fig8}. The relatively small contributions
from the neutron and core  participants are not included and the
longitudinal and transverse momentum distributions are therefore
identical.  The inset shows the distributions in the same unit for
different beam energies.}
\label{fig12}
\end{figure}

The momentum distributions of $^{9}$Li are not far from those found in
the simplest approximation described by the Fourier transform of the
initial three-body wave function. However, the neutron momentum
distributions are strongly influenced by the final state interaction.
Instead the momentum distribution of the center of mass of $^{10}$Li
shown in Fig. \ref{fig12} reveals direct information about the neutron
momentum distribution in the initial three-body system. The process is
removal (stripping or diffraction) of one halo neutron by the target
with the remaining $^{10}$Li system as the two spectators.  In the
center of mass of the three-body projectile the momentum distributions
of $^{10}$Li and the participant neutron are then identical.

This observable is insensitive to the final state interaction where
the opposite erroneously was postulated in \cite{aum98}. On the other
hand the sensitivity to the shadowing parameters is large
\cite{ale98,gar98}.  Experimental data could then be very useful to
check the validity of the shadowing parameters extracted in the
previous section from the comparison between computed momentum
distributions and available experimental data.  The final state
momentum of the participant neutron does not enter in the measured
momentum and the longitudinal and transverse $^{10}$Li momentum
distribution are therefore identical.  This can also be understood
from the fact that the initial three-body momentum of the participant
neutron does not have a preferred direction. The inset of
Fig. \ref{fig12} again shows the variation of the distribution with
the beam energy.  The behavior is the same as discussed in connection
with the previous figures. The related key numbers are given in Table
\ref{tab3}.

\begin{figure}[ht]
\centerline{\psfig{figure=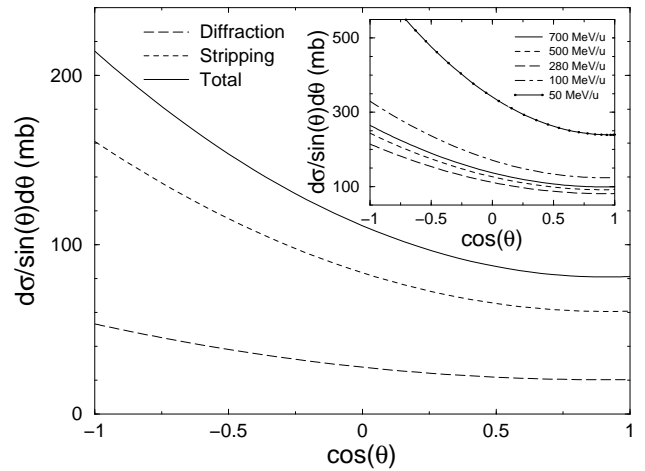,width=8.5cm,%
bbllx=2.7cm,bblly=1.7cm,bburx=18.8cm,bbury=24.0cm,angle=270}}
\vspace{0.2cm}
\caption[]{The distribution of the relative angle $\theta$ in the rest
frame of the initial three-body system. The angle $\theta$ is defined
as the angle between the center of mass momentum of $^{10}$Li and the
relative neutron-$^9$Li momentum. The reactions are the same as in
Fig. \ref{fig8}. The relatively small contributions from the neutron
and core participants are not included. The inset shows the
distributions for different beam energies. }
\label{fig13}
\end{figure}

The momentum distributions were recently supplemented by an angular
correlation measurement, where the observable is the angular
distribution of the relative momentum between the detected neutron and
the core in a coordinate system with the $z$-axis along the center of
mass momentum of the neutron-plus-core spectator system \cite{chu97}.
This observable is shown in Fig. \ref{fig13} for our standard case of
$^{11}$Li fragmentation on carbon.  In the computation we assumed that
$\theta$ is constructed as the angle between the momentum of the
center of mass of the $^{10}$Li spectator system and the relative
momentum between $^9$Li and the spectator neutron. The small
contributions, where the neutron or the core in $^{10}$Li are the
scattered participants, are not included in this figure, because they
are small and furthermore almost completely excluded in the experiment.
An estimate of the shape and size of one of the neglected
contributions can be found in \cite{gar98b}, where both $^6$He and
$^{11}$Li fragmentation are discussed.

The computations as always involve Eqs.(\ref{eq5}) and (\ref{eq6}).
The $s$-wave alone produces a constant angular distribution. The
$p$-wave alone produces a distribution of the form $1 + a \cos^2
\theta$, i.e. symmetric around $\cos \theta =0$. For $p$-waves with
the projection $m = \pm 1$ on the $^{10}$Li momentum, $a$ is very
small and negative, whereas $a$ for $m=0$ is positive and of the order
one. In other words the variation in the angular distribution is
almost entirely due to the $p$-wave with projection 0. If both $s$ and
$p$-waves are present the distribution takes the form $1 + a \cos^2
\theta + b \cos \theta$, i.e. becomes asymmetric due to mixing between
these partial waves.

Thus our pronounced asymmetry arises from the $s$ and $p$ mixing term,
which dominates the angular variation, since the largest contribution
from the $s$-wave alone is constant. A dominating $p$-wave would have
produced a much more symmetric distribution.  Substantial deviations
between measured and computed distributions would indicate a selective
reaction mechanism emphasizing specific partial waves.  The shape of
the angular distribution is independent of the beam energy as shown in
the inset of Fig. \ref{fig13}. Only the absolute values of the cross
section changes in accordance with Fig. \ref{fig2}. The shape in
Fig. \ref{fig13} deviates slightly from that of \cite{gar98b} due to
the higher $p$-wave content.

\begin{figure}[ht]
\centerline{\psfig{figure=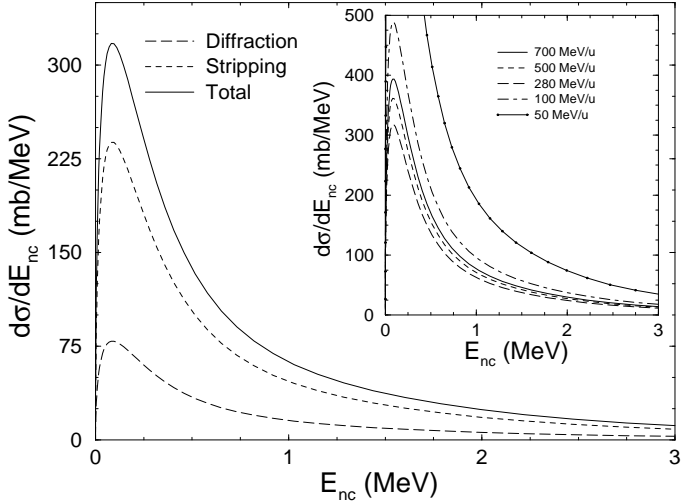,width=8.5cm,%
bbllx=2.8cm,bblly=2.8cm,bburx=18.8cm,bbury=23.4cm,angle=270}}
\vspace{0.2cm}
\caption[]{The invariant mass spectrum of $^{10}$Li for the cases in
Fig. \ref{fig8}. The rest mass is subtracted. The relatively small
contributions from neutron and core participants are not included. The
inset shows the distributions for different beam energies. }
\label{fig14}
\end{figure}

The momentum distributions reveal properties of the initial three-body
system and the reaction mechanism. The invariant mass spectrum of
$^{10}$Li after fragmentation of $^{11}$Li on a light target carry in
addition information about the properties of the two-body system. The
computed spectrum is shown in Fig. \ref{fig14}, where we only included
contributions from the spectator neutron, i.e. the invariant mass
$E_{nc}$ is constructed with the momentum of the spectator neutron
even when the participant neutron is scattered by the target.  In the
center of mass system of $^{10}$Li the invariant mass is, apart from
the rest mass, simply the total kinetic energy of the two-body system.

The spectrum must start from zero due to the phase space. The very
low-lying peak is a signature of a dominating $s$-wave
contribution. The height of the peak, not the position, reflects the
energy of the underlying two-body virtual $s$-state. In contrast, a
dominating $p$-wave contribution produces a peak at the energy of the
corresponding two-body resonance. In the present case the $p$-wave
contribution is smeared out in a region around 0.5 MeV. A narrow
$p$-wave resonance would show up as a peak in this invariant mass
spectrum \cite{gar97}. The measured spectrum then indicates a
low-lying $s$-state and a higher-lying $p$-resonance with a moderate
width, see Fig. \ref{fig4}c, where the curve computed with potential I
is the same as shown in Fig. \ref{fig14} for a beam energy of 280 MeV/u. The
contributions from stripping and diffraction as well as the energy
dependence shown in the inset is consistent with the results in
Fig. \ref{fig2}. The energy dependence of the peak heights and the
widths are given in Table \ref{tab3}.

As seen in Table \ref{tab3} all the cross sections first decrease with
energy and then after a minimum slowly increase again. When the
contribution from the scattered neutron is included the widths follow
qualitatively the same pattern, although much less pronounced, and in
particular they essentially do not vary for energies above around 200
MeV/u. When only the spectator neutron contributes the widths are
energy independent. This behavior disagrees with that of \cite{zhu95},
but is on the other hand consistent with the experimental results
described in \cite{orr97,gei97}.

In this comparison we implied that our FWHM is the same quantity as
the width (or $\Gamma$-values) discussed in previous experimental
papers. Although these quantities are strongly correlated this is not
strictly true, since the FWHM precisely is defined as the full width
at half maximum of the computed distribution whereas the experimental
width is obtained as the width parameter in a fitted function. This
easily shifts the emphasis from the small momenta in the calculations
to the large momenta of the tail in the experimental analysis. This
could easily produce uncertainties of several MeV in the FWHM.

As we have demonstrated the distributions are more complicated than
simple one-parameter functions. Comparison of our FWHM with published
experimental widths could then be rather misleading. The only safe
procedure is to compare directly with the experimental data as we did
throughout this paper. On the other hand then we immediately face the
problem that the data too often include purely experimental effects
related to beam profile, acceptance range, resolution and target
thickness. It is therefore important to compare with the properly
interpreted data or alternatively to fold the experimental effects
with the calculated results.

\section{Discussion and conclusions}

Fragmentation reactions of weakly bound two-neutron halo nuclei
provide detailed information about structure and reaction mechanism of
corresponding three-body systems. The three-body structure can be
computed to the needed accuracy provided the interactions are
specified whereas the reactions are much more difficult to treat
propertly.  Our aim is to understand and describe the principal
features of all fragmentation observables qualitatively and
quantitatively to an accuracy comparable with that of the experimental
data. As an important step we have chosen to investigate $^{11}$Li
considered as a three-body system in a physically simple and
transparent model based on geometric properties and phenomenological
interactions.

One main assumption is often the sudden approximation where the
reaction is instantaneous. The reaction time must then be short
compared to the intrinsic time scale of the relative motion of both
the three halo particles and the nucleons in the target and the
core. Thus the intrinsic motion must be frozen during the collision or
equivalently the beam energy must be large compared to the three-body
binding energy and the Fermi energy of the target and the core. As
expected this approximation has successfully passed the tests at high
energies.

In the present model we do not directly use the sudden
approximation. We fully include the interaction between the target and
each of the halo particles by use of a phenomenological optical
model. The description of the interactions is then only limited by the
validity range of the parameters employed in the optical model. On the
other hand then only elastic scattering is described in details while
all other processes are included as absorption from the elastic
channel. Fortunately this is precisely the level of information
required in discussions of fragmentation reactions, because the
inelastic channels overwhelmingly produce different reaction products
or particles scattered outside the detection range in the forward
direction.

Furthermore we only include the interaction between one halo particle
(participant) and the target while neglecting the interactions between
the other two halo particles (spectators) and the target. In addition
we also neglect the interactions between the spectators and the
participant-target system in the final state. The halo particles must
then interact independently with the target as three spatially
correlated but non-interacting clusters of nucleons. The motion of the
spectators must remain undisturbed by the participant-target
interaction. More precisely the two criteria for the validity of the
model are that (i) the sum of the participant and target radii is less
than the halo radius and (ii) the intrinsic velocity of the partipant
within the halo is much smaller than the velocity of the projectile or
perhaps better that the characteristic time for the intrinsic halo
motion is much larger than the collision time. 

These main approximations are justified for weakly bound and spatially
extended halos colliding with a target with an energy per nucleon
larger than the intrinsic kinetic energy of the participant within the
halo, i.e. roughly the usual Fermi energy for nuclei although in
principle unrelated to the Fermi energies of the nucleons within
target and core. With these approximations one halo particle can
interact with the target without disturbing the motion of the other
two. The total cross section is obtained by adding the contributions
from the three possible participants. The sudden approximation is
reached when the optical model is reduced to a black sphere when the
elastic scattering also is neglected.

The model is described for point particles and the necessary
generalization to finite radii involves the concept of shadowing. We
eliminate the geometric configurations in the three-body wave function
where the spectators move in the shadow of the participant. The need
for this correction arises from the simplifying choice of treating the
participant-target interaction properly while leaving the spectators
untouched in their initial state. If the spectators are close to the
participant they would be either absorped or similarly
scattered. However, these events contribute in the model computations
with the probability given in the initial wave function. Consequently
we must omit those unwanted configurations. Another improved treatment
with better final state wave functions could directly take these
effects into account.

To use the model we must specify the interactions and the shadowing
parameters.  Within the halo projectile we have the neutron-neutron
and neutron-core interactions supplemented by the three-body
force. They are parametrized as gaussians to reproduce the $^{11}$Li
binding energy, give a neutron-$^{9}$Li $p_{1/2}$-resonance at 0.5 MeV
with a width of 0.4 MeV and finally to produce about 40\% of
$p^2$-configurations in the $^{11}$Li wave function. These
requirements are necessary to reproduce various experimental data. The
only freedom left for the halo interactions is then the
spin-splitting, arising from the two different couplings to the core
spin of $3/2$, of both the $s_{1/2}$ and $p_{1/2}$-states in the
neutron-core system. The related spin-splitting parameters influence
neither the above data nor the fragmentation data. Good agreement with
the data then indicates approximately correct statistically averaged
positions of the $s$ and $p$-states in $^{10}$Li, i.e. 0.71 MeV and
0.76 MeV.

The two-body interactions between halo particles and target are
described by use of the phenomenological optical model with parameters
adjusted to reproduce the corresponding elastic scattering and
absorption cross section data. The two shadowing parameters related to
the sizes of halo particles and target are determined to reproduce
both the absolute two-neutron removal cross section and the momentum
distributions after fragmentation.

We compute all momentum distributions related to fragmentation of
$^{11}$Li on carbon. For the same reactions we also compute the
invariant mass of $^{11}$Li and the neutron angular distribution,
which recently was measured for $^{6}$He fragmentation. These
observables are in general consistent with the available measurements.
When the neutron and core participants are scattered they receive
momentum tranfer perpendicular to the beam and the transverse momentum
distributions are therefore broader than the corresponding longitudinal
distributions. This is then a direct effect of the diffraction
mechanism.

The distribution for neutrons is relatively narrow due to the final
state interaction, which affects the core less. The distribution for
the center of mass of the core-plus-neutron spectators is the broadest
reflecting the extension of the initial wave function. Increasing the
shadowing parameters decrease the widths of the distributions. The
invariant mass reveals information about the low-lying continuum
structure of the neutron-plus-core system. The large and very
low-lying peak is the signature of a low-lying virtual $s$-state while
the shoulder indicates a low-lying and fairly broad $p$-resonance. The
angular correlation of the emitted neutrons in the neutron-plus-core
center of mass system is highly asymmetric revealing that $s$ and $p$
relative neutron-core states roughly are equally populated in
$^{11}$Li.

The experimental data are available for several energies and targets,
but systematic high precision data given as function of energy for one
target does not exist at the moment. We can compute absolute values of
a number of differential cross sections.  However, in this paper we
confined ourselves to the energy dependence of three-body observables
for fragmentation on a carbon target with one excursion to an aluminum
target. The distributions are essentially independent of target, but
the absolute differential cross sections increase with target
size. The scaling with target size seems to be somewhat larger than
the square of the target radius.

The energy dependence for the given carbon target is computed for a
$^{11}$Li beam of 50 to 900 MeV/u. The widths of the distributions are
essentially constant above around 200 MeV/u and slightly increasing
towards lower energies. The absolute values follow the
participant-target cross sections. For neutrons this means a smooth
increase with energy above around 150 MeV/u and a strong increasee
towards lower energies. Furthermore diffraction contributes much less
than absorption at energies above 200 MeV/u whereas the inverse is
true for energies below 50 MeV/u. This has the consequence that the
widths of the transverse distributions are broader at low energies due
to the domination of diffraction.

These predictions presupposes that the model is valid at the low
energies and the energy dependence of the parameters are correctly
included. The criteria for validity indicated relative neutron-target
energies above around 20 to 30 MeV and perhaps even lower for very
pronounced halo systems. The optical model parameters for the
neutron-carbon potential are adjusted to scattering data down to these
energies, but the shadowing parameters are assumed to be constant.
These parameters have a strong influence on the absolute values of the
cross sections and a significant, but much less pronounced, influence
on the shapes of the distributions. The predicted widths could perhaps
be systematically changed by small amounts due to such possible
energy dependence.

In conclusion, we computed systematically essentially all observables
for the $^{11}$Li three-body fragmentation on a carbon target. The
same consistent model is used throughout. Most of the computations are
in agreement with available measurements. This strongly indicates that
the reaction mechanism essentially is correctly described in the
model. The predictions are therefore useful as the unit for comparison
with future experimental data. Detection of discrepancies would then
be significant and therefore also very suggestive of necessary
improvements like, for eaxmple, different dependencies of some of the
parameters. The same consistent model for all observables is crucial
at the present level of accuracy and understanding. In this connection
it is worth keeping in mind that treating $^{11}$Li as a three-body
system is an approxomation and the intrinsic structure must be
unavoidable at some point in the interpretation.

\appendix
\section{Spin 1/2 scattering on spin zero targets}

In Eq. (\ref{eq2}) we give the transition amplitude for the scattering process
between the participant particle $i$ and the target. Assuming that
particle $i$ has spin $s_i=1/2$ and the target has spin zero, we can
write the transition amplitude as \cite{sit71}.

\begin{equation}
T_{\Sigma_i,\Sigma^\prime_i}^{(0i)} = -\frac{2\pi}{\mu_{0i}}
\langle \chi_{s_i \Sigma_i^\prime} | g(\theta) + h(\theta) 
(\bd{n} \cdot \bd{\sigma}) | \chi_{s_i \Sigma_i} \rangle \; ,
\end{equation}
where $\theta$ is the angle between $\bd{p}_{0i}$ and
$\bd{p}^\prime_{0i}$, $\bd{\sigma}$ are the Pauli spin matrices and
the functions $g(\theta)$ and $h(\theta)$ are given by
Eqs. (\ref{gfu}) and (\ref{hfu}), respectively. The vector $\bd{n}$ is
defined as
\begin{equation}
\bd{n}=\frac{\bd{p}_{0i} \times \bd{p}_{0i}^\prime}
        { |\bd{p}_{0i} \times \bd{p}_{0i}^\prime| } \;,
\end{equation}
\begin{equation}
(\bd{n} \cdot \bd{\sigma} ) = (\bd{n} \cdot \bd{\sigma} )^\dagger=
\left( \begin{array}{cc}
        n_z & n_x+in_y \\
        n_x-in_y & -n_z \\
       \end{array} \right) \; ,
\label{nsig}
\end{equation}
which implies that $(\bd{n} \cdot \bd{\sigma} )^2=1$. 

We can then rewrite the key part of Eq. (\ref{e7}) as
\begin{eqnarray}
\lefteqn{
\sum_{M s_{jk} \Sigma_{jk} \Sigma^\prime_i} \left|
\sum_{\Sigma_i} T^{(0i)}_{\Sigma_i \Sigma^\prime_i}
M_{s_{jk} \Sigma_{jk} \Sigma_i}^{JM}
\right|^2 = }  \nonumber \\ &&
\frac{(2\pi)^2}{\mu_{0i}^2}
\sum_{\Sigma^\prime_i \Sigma_i \Sigma^{\prime \prime}_i}
\langle \chi_{s_i \Sigma_i} | g^*(\theta) + h^*(\theta)
(\bd{n} \cdot \bd{\sigma}) | \chi_{s_i \Sigma_i^\prime} \rangle \nonumber \\ &&
\times \langle \chi_{s_i \Sigma_i^\prime} | g(\theta) + h(\theta)
(\bd{n} \cdot \bd{\sigma}) | \chi_{s_i \Sigma^{\prime \prime}_i} \rangle
 \nonumber \\ && \times 
\sum_{M s_{jk} \Sigma_{jk}} M_{s_{jk} \Sigma_{jk} \Sigma_i}^{JM*}
M_{s_{jk} \Sigma_{jk} \Sigma_i^{\prime\prime}}^{JM} =
 \nonumber \\ &&
\frac{(2\pi)^2}{\mu_{0i}^2}
\sum_{\Sigma_i \Sigma^{\prime \prime}_i}
\langle \chi_{s_i \Sigma_i} | |g(\theta)|^2 + |h(\theta)|^2 +
2 \Re[g(\theta) h^*(\theta)] \nonumber \\ && 
\times (\bd{n} \cdot \bd{\sigma} ) 
| \chi_{s_i \Sigma^{\prime \prime}_i} \rangle
\sum_{M s_{jk} \Sigma_{jk}} M_{s_{jk} \Sigma_{jk} \Sigma_i}^{JM*}
M_{s_{jk} \Sigma_{jk} \Sigma_i^{\prime\prime}}^{JM} =
 \nonumber \\ &&
\frac{(2\pi)^2}{\mu_{0i}^2} (|g(\theta)|^2 + |h(\theta)|^2)
\sum_{M s_{jk} \Sigma_{jm} \Sigma_i} 
|M_{s_{jk} \Sigma_{jk} \Sigma_i}^{JM}|^2 \; ,  \label{A4}
\end{eqnarray}
where we used that
\begin{eqnarray}
\sum_{\Sigma_i \Sigma^{\prime \prime}_i}
\langle \chi_{s_i \Sigma_i} | (\bd{n} \cdot \bd{\sigma} )
| \chi_{s_i \Sigma^{\prime \prime}_i} \rangle \nonumber \\ 
\times \sum_{M s_{jk} \Sigma_{jk}} M_{s_{jk} \Sigma_{jk} \Sigma_i}^{JM*}
M_{s_{jk} \Sigma_{jk} \Sigma_i^{\prime\prime}}^{JM} = 0
\label{rem}
\end{eqnarray}
as seen from Eq. (\ref{nsig}) and the fact that the matrix
\begin{equation}
B^J_{\Sigma_i, \Sigma^\prime_i} = 
\sum_{M s_{jk} \Sigma_{jk}} M_{s_{jk} \Sigma_{jk} \Sigma_i}^{JM*}
M_{s_{jk} \Sigma_{jk} \Sigma_i^\prime}^{JM}
\label{bmat}
\end{equation}
is diagonal with identical diagonal elements \cite{gar97}.
Insertion of Eq. (\ref{A4}) into Eq. (\ref{e7}) and use of
Eqs. (\ref{pha}), (\ref{sudden}) and (\ref{el2}) then immediately lead to
Eq. (\ref{eq5}) and (\ref{el1}).

{\bf Acknowledgments.} We thank K. Riisager for continuous discussions
and many suggestions.


\end{document}